\documentstyle[pra,aps,psfig,epsfig,psfrag]{revtex}

\def \lta {\mathrel{\vcenter{\hbox{$<$}\nointerlineskip\hbox{$\sim$}}}}
\def \gta {\mathrel{\vcenter{\hbox{$>$}\nointerlineskip\hbox{$\sim$}}}}
\begin{document}
\draft

\title{High-energy neutrino conversion and the lepton asymmetry in the universe}

\author{C.Lunardini$^{a)}$, A.Yu.Smirnov$^{b)}$}

\bibliographystyle{unsrt}

\address{ a) SISSA-ISAS, via Beirut 2-4, 34100 Trieste, Italy \\
 and INFN, sezione di Trieste, via Valerio 2, 34127 Trieste, Italy\\
\vspace{0.2cm}
b) The Abdus Salam ICTP, Strada Costiera 11, 34100 Trieste,
Italy \\
and Institute for Nuclear Research, RAS, Moscow, Russia}

\maketitle

\begin{abstract}
\noindent
We study matter effects on oscillations of high-energy neutrinos in the Universe. Substantial effect can be produced by scattering of the neutrinos from cosmological sources ($z\gta 1$) on the relic neutrino background, provided that the latter has large CP-asymmetry: $\eta\equiv (n_\nu-n_{\bar{\nu}})/n_\gamma\gta 1$, where $n_\nu$, $n_{\bar{\nu}}$ and $n_\gamma$ are the concentrations of neutrinos, antineutrinos and photons.  We consider in details the dynamics of conversion in the expanding neutrino background. Applications are given to the diffuse fluxes of neutrinos from GRBs, AGN, and the decay of super-heavy relics.  We find that the vacuum oscillation probability can be modified by $\sim (10-20)\%$ and in extreme cases allowed by present bounds on $\eta$ the effect can reach $\sim 100\%$.
Signatures  of matter effects  would consist 
(i) for both active-active and active-sterile conversion, in a deviation  of the numbers of events produced in a detector by neutrinos of different flavours, $N_{\alpha}~(\alpha=e,\mu,\tau)$, and of their ratios from the values given by vacuum oscillations; such deviations can reach $\sim 5-15\%$, (ii) for active-sterile conversion, in a
characteristic energy dependence of the ratios $N_{e}/N_{\mu},N_{e}/N_{\tau},N_{\mu}/N_{\tau}$.
Searches for these matter effects will probe large
CP and lepton asymmetries in the universe.

\end{abstract}


\section{Introduction}
\label{sec:1}
The detection of high-energy cosmic neutrinos and detailed studies of their properties are among the main challenges in astrophysics and cosmology. 
This will give unique information about the structure of the universe, mechanisms of particle acceleration, sources of cosmic rays, properties of the galactic and intergalactic media.  They will have also important implications for neutrino properties (masses, mixings, etc.) and for particle physics in general. 
 
Intense fluxes of neutrinos, with energies up to 
$\sim 10^{21}~{\rm eV}$, are supposed to be produced by cosmological objects 
 like  Gamma Ray Bursters (GRBs) and Active Galactic Nuclei (AGN)  
\cite{rewgrb}.  
It was suggested  that neutrinos of  energies as high as 
$10^{22}-10^{24}~{\rm eV}$ could be
produced by topological defects like cosmic strings, necklaces  and domain 
walls \cite{Bhattacharjee:1997ps}. Furthermore, neutrinos produced by the decay of super-heavy
particles
have been
considered in connection to the problem of ultra-high energy cosmic rays
exceeding the GZK cutoff \cite{Weiler:1997sh}.

The detection of high-energy neutrinos from cosmological sources is 
challenging for  neutrino telescopes.  The existing large water,  ice, or airshower 
experiments \cite{Halzen:1998mb_s}  open some possibility of detection; 
however a detailed study requires larger detectors to be realized in future.

The properties of high-energy neutrino fluxes can be modified by oscillations on the way from the sources to the Earth.  In particular it was marked \cite{Learned:1995wg} that oscillations lead to the appearance of tau neutrinos in the high-energy neutrino flux. Moreover, the study of oscillation effects opens 
the possibility to probe neutrino mixings and distinguish between different mass spectra  \cite{Bento:1999bb}.  In all these studies  vacuum oscillations have been considered only.   
In this connection, we address here two questions:

\begin{enumerate}
\item 
Are matter effects important for high-energy neutrinos propagating in the universe?

\item
 Which information on the properties of the interstellar and intergalactic medium can be obtained from the study of these effects? 

\end{enumerate}

During their travel from the production
point to the detector, the neutrinos cross large amounts of matter, which could
induce significant refraction and conversion.
In ref. \cite{Lunardini:2000sw} we considered 
the interaction of neutrinos with the matter of the source  for neutrinos produced in
GRBs and AGN. The effects of matter on vacuum oscillations appeared to be 
small. It was also found \cite{Lunardini:2000sw} that the neutrino-neutrino interaction in the dark matter  
 halos of galaxies does not affect the vacuum oscillations significantly.  
Conversely, strong matter effects are
not excluded for neutrinos crossing media of larger size, like the halos
of clusters of galaxies.  Furthermore, neutrinos from cosmological sources travel for so large distances in the intergalactic space that the universe itself, with
its particle content, can be considered as a medium producing refraction 
effects. 
In \cite{Lunardini:2000sw} we found that  significant
 conversion can be realized for neutrinos crossing cosmological distances in the universe with strongly CP-asymmetric neutrino background.
\\

In this paper we analyze this possibility in detail.  We discuss the refraction and conversion effects of the background on high-energy neutrinos from cosmological sources and on neutrinos of the background itself.
 
Let us  describe the relic neutrino gas  by the number densities of the various
flavours, $n_\alpha$ ($\alpha=e, \bar{e}, \mu, $ etc.), and by the
CP-asymmetry $\eta_{\nu}$ defined as:
\begin{eqnarray} 
\eta_{\nu}\equiv (n_\alpha-n_{\bar{\alpha}})/n_\gamma~,
\label{eq:asym}
\end{eqnarray}
where $n_\gamma$ is the  concentration of photons.
 
The Big Bang 
Nucleosynthesis (BBN) and structure formation \cite{Kang:1992xa,Sarkar}  
admit large CP-asymmetries
for muon and tau neutrinos, while the  asymmetry for the
electron neutrino is strongly constrained:
\begin{eqnarray} 
|\eta_{\mu,\tau}|\lta 10~, \hskip 1.5cm 
  -0.01\lta \eta_{e}\lta 0.3~.
\label{eq:bbnbound}
\end{eqnarray}
Large asymmetries have also important implications on the properties of the spectrum of the cosmic microwave background 
radiation (CMBR) \cite{Sarkar}. The recent results on the second acoustic peak of the CMBR from BOOMERANG and MAXIMA-1 experiments \cite{deBernardis:2000gy}
seem to favour a large lepton asymmetry,  $\eta_\nu\sim 1$ \cite{Lesgourgues:2000eq,Hannestad:2000hc,Orito:2000zb}. 
In particular, a satisfactory interpretation of the data requires \cite{Hannestad:2000hc}: 
\begin{eqnarray} 
|\eta_{e,\mu,\tau}|\lta 2.2~,
\label{eq:cmbrbound}
\end{eqnarray}
thus providing a stronger restriction of the allowed $\eta_\mu$ and $\eta_\tau$ values with respect to (\ref{eq:bbnbound}).
In our discussion we will consider asymmetries $\eta_\mu$ and $\eta_\tau$ as large as $|\eta_{\mu,\tau}|\sim 1-2$, according to the upper limit (\ref{eq:cmbrbound}); however results will be given also for larger values, allowed by the less stringent bound (\ref{eq:bbnbound}).
   
We want to underline here that the realization of large CP-asymmetries in the individual lepton flavours  is consistent with zero lepton asymmetry. This corresponds to: 
\begin{eqnarray}
\eta_e+\eta_\mu+\eta_\tau=0~,
\label{eq:leptsym}
\end{eqnarray} 
that is, to zero total lepton number.
Large lepton asymmetry, in contrast, implies large CP-asymmetry.

The paper is organized as follows.
In section \ref{sec:0} the properties of the relic neutrino background are discussed.  We show that significant matter effects on high-energy neutrinos require large CP-asymmetry of the background, and study the refraction and conversion effects in the background itself. 
In sections \ref{sec:new} and \ref{sec:3} we study matter effects on high energy neutrinos produced at cosmological distances.  Applications are given  in section  \ref{sec:4} to the diffuse fluxes of neutrinos from AGN, GRBs and from the decay of heavy relics.  In section \ref{subsec:4.3} we  discuss the experimental signatures of matter effects.
Conclusions follow in section \ref{sec:5}.

\section{Across the Universe}
\label{sec:0}

Let us consider the interactions of high-energy neutrinos propagating from cosmological sources to the Earth. 
These neutrinos cross layers of matter in the source itself, then interact with particles in the interstellar and intergalactic media, and finally interact in the matter of our cluster of galaxies and of our galaxy.   

In what follows we will discuss interactions in the intergalactic medium.  The effects of the matter of the sources can be neglected  \cite{Lunardini:2000sw}. As we will show later, for neutrino oscillation parameters and energies relevant for this discussion also the effect of the galactic halo and of the halo of the cluster of galaxies are very small  \cite{Lunardini:2000sw}.

\subsection{Minimum width condition}

The necessary condition for significant matter effect is the minimum width condition \cite{Lunardini:2000sw}.  
We define the width of the medium  as  the integrated concentration of the particle background of  the universe along the  path travelled
by the neutrino beam:
\begin{equation}
d\equiv\int^{t_0}_t n(t') dt'~.
\label{eq:ddef} 
\end{equation} 
Here $t$ is the epoch of production of the neutrinos and $t_{0}\sim 10^{18}$ s  is the present epoch.
Using the scaling of the concentration $n(t)\propto t^{-2}$,
one finds:
\begin{eqnarray}
&&d(t)=d_{U} \left[{t_0\over t} -1\right]\simeq d_U \left[(1+z)^{3 \over 2}-1\right]~,
\label{eq:dtev} \\
&&d_U\equiv t_0 n_0~,
\label{eq:du}
\end{eqnarray}
where $n_0$ is the present concentration of the background and we have introduced the redshift $z\equiv\left({t_0/ t}\right)^{2/3}-1$.

The minimum width condition can be written as \cite{Lunardini:2000sw}:
\begin{eqnarray}
r\equiv {d\over d_0}\geq 1~,
\label{eq:rmin}
\end{eqnarray}
where $d_0$ is the refraction width of the medium:
\begin{equation} 
d_0\equiv{\pi \over 2}{n \over V}~,
\label{eq:d0}
\end{equation}
with $V$ being the effective matter potential in a given neutrino conversion channel.  The width $d_0$ corresponds to the distance in matter at which the oscillation phase induced by matter equals $\pi/2$. Notice that, since $V\propto n$, $d_0$ does not depend on the density of the medium and is determined by the properties of the interaction.  The usual weak interaction gives $d_0\sim 1/G_F$, where $G_F$ is the Fermi constant.    
\\

\noindent
Let us consider the fulfillment of the condition (\ref{eq:rmin}) for different components of the intergalactic medium.
\\

\noindent
1). Due to the very small concentration of nucleons and electrons, the width of these components is extremely small, $d_B/d_0\ll 1$, even for neutrinos produced at cosmological distances.  Indeed, the baryon concentration can be estimated as $n_B=n_\gamma \eta_B$  where $\eta_B=10^{-10}-10^{-9}$ is the baryon asymmetry of the universe and  $n_\gamma$ the photon concentration. At present   time  $n_\gamma=n^0_\gamma\simeq 412~{\rm cm^{-3}}$.  Taking, e.g., production epoch $z\simeq 1$  we find from eqs. (\ref{eq:dtev})-(\ref{eq:d0}) that for baryons $r\equiv r_B\sim 10^{-11}$. 
\\

\noindent
2). The scattering on the electromagnetic background has negligible effect due to the smallness of interaction. The neutrino-photon potential is of the second  order in the Fermi constant and depends on the energy of the neutrino beam and on the temperature and concentration of the photon gas \cite{Levine:1967nc,Dicus:1997rw}. 
Using the results of ref. \cite{Dicus:1997rw} (see also the discussion in \cite{raf})  we find   from eqs. (\ref{eq:dtev})-(\ref{eq:d0}) $r_\gamma \lta 10^{-8}$ for neutrino energy $E\lta 10^{21}$ eV and production epoch $z\simeq 1$.  
\\

\noindent
3). The effect of the scattering on the neutrino background can produce significant effect if the background has large CP-asymmetry\footnote{For CP-symmetric background significant effects can appear at large temperature, $T\gta 1$ MeV, due to thermal effects \cite{Notzold:1988ik}, or at extremely high neutrino energies, due to  neutrino-antineutrino scattering in the resonant $Z^0$ channel \cite{Lunardini:2000sw}.}. Indeed, for asymmetry $\eta_\nu\sim 1$ we get  $d_\nu\sim d_0$. Clearly,  
if the lepton asymmetry is of the order  of the baryon one,  $\eta_\nu\simeq \eta_B$,  the width is negligibly small: $d_\nu\sim d_B\ll d_0$. 
\\

Let  us consider the minimum width condition for neutrino background in more detail. 
The effective potential due to the scattering of neutrinos on the relic neutrino background can be written as 
\begin{eqnarray}
&&V=F \eta_\nu  \sqrt{2} G_{F} n_\gamma~, 
\label{eq:potefunivb} 
\end{eqnarray}
where $F$ is a constant of order $1$ which depends on the specific conversion channel (see sections \ref{subsec:new.1} and \ref{subsec:3.1}). 
With the potential  (\ref{eq:potefunivb}), eqs. (\ref{eq:dtev}), (\ref{eq:rmin}) and (\ref{eq:d0}) give the condition:
\begin{eqnarray}
r(z)=1.6\cdot 10^{-2} |F| \eta_\nu \left[(z+1)^{3\over
2}-1\right]\geq 1~.
\label{eq:dasym} 
\end{eqnarray}
For neutrinos produced in the present epoch, $z\simeq 0$ and values of  $\eta_\nu$ allowed by the bounds (\ref{eq:bbnbound})-(\ref{eq:cmbrbound}) the
condition (\ref{eq:dasym}) is not satisfied:  $r(z\sim 0)\ll 1$.
From (\ref{eq:dasym}) we can define
 the epoch $z_d$ which corresponds to
$r=1$:
\begin{eqnarray}
1+z_d=\left[1+{1 \over 1.6\cdot 10^{-2} |F| \eta_{\nu}}\right]^{2\over 3}~, 
\label{eq:zd} 
\end{eqnarray}
so that for neutrinos produced at $z\geq z_d$ the minimum width condition is fulfilled.
Taking for instance $\eta_{\nu}=1$ and $F=2$ we get $z_d\simeq 9$; by requiring $r\simeq 0.3$ (which corresponds to $10\%$ matter effect \cite{Lunardini:2000sw}) we find  $z_d\simeq 3.7$.  
\\

The following remark is in order.
The condition (\ref{eq:rmin}) is necessary but not sufficient to
have significant matter effects. In particular, for the case of oscillations in 
uniform medium and  small mixing angles ($\sin 2\theta \lta 0.3$) we have found in ref. \cite{Lunardini:2000sw}
that the width  $d_{min}$
needed to have conversion probability larger than $1/2$ equals:
\begin{eqnarray}
 d_{min}={d_0 \over \tan 2\theta}>d_0~.
\label{eq:dmin} 
\end{eqnarray} 
This quantity represents an absolute minimum.  
For media with varying density the required width  is larger
than $d_{min}$ \cite{Lunardini:2000sw}. 
\\

We conclude, then, that the only component of the intergalactic medium which can produce a significant matter effect is a strongly CP-asymmetric neutrino background, with $\eta_\nu \gta 1$. Moreover, cosmological epochs of neutrino production are required: $z\gta 3$.


\subsection{Properties of the relic neutrino background}
\label{sec:2}
Neutrino mixing and oscillations modify the flavour composition of the neutrino background, so that one expects the present values of the  CP-asymmetries in the various flavours
 to be different from those at the epoch of BBN; the latter are constrained by the bounds (\ref{eq:bbnbound}).  
 
In this section we assume that large CP-asymmetries
are produced at some epoch before the
BBN, i.e., at temperature $T\gta T_{BBN}\simeq 1$ MeV, and study how they evolve with time. 
The evolution of the flavour densities $n_e$, $n_\mu$ and $n_\tau$ is  a non-linear many-body problem, which, in general, requires a numerical treatment\cite{Samuel:1993uw,Sigl:1993fn}.
In some specific cases, however, an analytical description is possible\cite{Kostelecky:1995dt} and conclusions can be obtained on general grounds.

\subsubsection{Three-neutrino system evolution}
\label{subsubsec:2.2.1}
Let us first consider the case of mixing between three active neutrinos, $\nu_e$, $\nu_\mu$, $\nu_\tau$.    
In refs.\cite{Samuel:1993uw,Kostelecky:1995dt} it has been shown  that the evolution of the flavour densities has peculiar aspects for the ideal case of a monoenergetic gas of neutrinos (with no antineutrinos, $n_{\bar \nu}=0$) initially produced in flavour states. For this specific ensemble of neutrinos the potential due to neutrino-neutrino interaction cancels in the evolution equation, so that the collective behaviour of the system is described by vacuum oscillations. 
This result holds with a good approximation \cite{Samuel:1993uw} also for realistic neutrino energy spectra and in presence of a small component of antineutrinos.  For this reason it can be applied to 
 our case of interest, in which the  background is strongly CP-asymmetric, $n_\nu\gg n_{\bar \nu}$, and neutrinos have a thermal spectrum.   In what follows we approximate the neutrino energies with the average thermal energy of the gas: $E\simeq <E_\nu>=\alpha T_\nu$, where  $T_\nu$ denotes the temperature of the neutrino gas. 
The numerical factor $\alpha$ depends on the CP-asymmetry of the background: we have $\alpha\simeq 3.15$ in absence of asymmetry, $\eta\simeq 0$, and $\alpha\simeq 3.78$ for $\eta\simeq 1$.
\\

The length scale of flavour conversion is given by the vacuum oscillation length:
\begin{eqnarray} 
l_v&=&{4 \pi E \over \Delta m^2}\simeq{4 \pi \alpha \beta T \over \Delta m^2}\nonumber \\
&=& 2.48 \cdot 10^{5}~{\rm cm}~ \alpha \beta \left({T\over {\rm 1 MeV}} \right) \left({{10^{-3}\rm eV^2}\over \Delta m^2} \right)~, 
\label{eq:lv}
\end{eqnarray}
where $\beta$ is the ratio between the temperature  of the neutrino background and the temperature $T$ of the electromagnetic radiation: $\beta\equiv T_{\nu}/T$. We have  $\beta=1$ before the electron-positron recombination epoch, $T\gta  0.5~{\rm MeV}$, and $\beta=(4/11)^{1/3}$ after this epoch.  
\\

\noindent
Besides oscillations, for temperatures $T\gta 1~{\rm MeV}$  other phenomena, and therefore other length scales, are relevant: 
\\

\begin{itemize}  
\item
Inelastic collisions.  Let us consider a system of two mixed neutrinos, $\nu_a$, $\nu_b$ ($a,b=e,\nu,\tau$). After its production as a flavour state, e.g. $\nu=\nu_a$, a neutrino oscillates in vacuum until a collision occurs with a particle $X$ of the background. At the time of the collision the quantum state of the neutrino is a coherent mixture of the two flavours:  $\nu=A \nu_a+ B \nu_b$. The effects of the collision depend on the specific reactions that take place \cite{Stodolsky:1987dx,Raffelt:1993uj,Smirnov:1987vz} (see also the discussion in \cite{raf}).  If the reaction is a scattering, $\nu X\rightarrow \nu X$, and the interaction is flavour blind, i.e. it is the same for the two flavours $a$ and $b$,  after the collision the neutrino continues to propagate as a coherent superposition of $\nu_a$ and $\nu_b$ and the collision does not affect oscillations.
For scattering with flavour-sensitive interaction or for absorption processes\footnote{In the situation we are considering the neutrinos are in thermodynamical equilibrium, so that their disappearance through a given reaction is balanced by their production through the inverse process.}, $\nu X\rightarrow any$,  the effect of the collision is to break the coherence between  $\nu_a$ and $\nu_b$, so that after the collision the two flavours evolve independently, developing vacuum oscillations until the next collision happens.   

As a result, one easily obtains that  for a beam of neutrinos propagating in a medium oscillations are damped according to the expression:   
\begin{eqnarray} 
n_a(L)={1 \over 2}+ \left(n^0_a-{1 \over 2} \right)\exp\left[{- {L \over l_c} \ln \left({1\over {1-2 P_c}} \right)}\right]~,
\label{eq:damp}
\end{eqnarray}
where $n^0_a$ and $n_a(L)$ are the fractions of $\nu_a$ in the neutrino beam at the production time and at distance $L$ from the production point.  Here $l_c$ is the coherence length, which represents the distance between two collisions, and  $P_c$ is the vacuum oscillation probability between two collisions: $P_c\simeq \sin^2 2\theta \sin^2(\pi l_c/l_v)$. 
From eq. (\ref{eq:damp}) we see that, if $P_c\neq 0$, with the increase of $L$ (i.e. of the number of collisions) the interplay of oscillations and collisions leads to the equilibration of the flavour densities: $n_a(L\rightarrow \infty)=n_b(L\rightarrow \infty)=1/2$. The convergence to this limit is determined by the equilibration length $l_{eq}\equiv l_c/\ln\left({1/ {|1-2P_c|}} \right)$.  
For small conversion probability, $P_c\ll 1$, the length $l_{eq}$ is much larger than $l_c$: $l_{eq}\simeq l_c/(2P_c)\gg l_c$. This is the case if the vacuum mixing is small and/or  the collisions are much more efficient than oscillations, $l_c\ll l_v$, so that the vacuum oscillation phase $\Phi=2 \pi l_c/l_v$ is small.  Thus, equilibration of the flavour densities can be obtained only after a large number of collisions: $n_{coll}=L/l_c\gta 1/(2P_c)$.  
Conversely, if $P_c\sim 1$ equilibration is achieved rapidly after few collisions: $n_{coll}\gta L/l_c$. This circumstance is realized if $l_c\gta l_v$ and the mixing is large, $\sin^2 2\theta \sim 1$.

Let us consider the coherence length, $l_c$, in more detail.
According to refs. \cite{Stodolsky:1987dx,Raffelt:1993uj}  $l_c$ can be written as: 
\begin{eqnarray} 
l_c(a,b)^{-1}={1 \over 2}\left[ \Gamma^{abs}(a)+\Gamma^{abs}(b)+\Gamma^{f s}(a,b) \right]~,
\label{eq:15}
\end{eqnarray}
where $\Gamma^{abs}(x)$ is the rate of absorption processes for the neutrino of flavour $x$ and $\Gamma^{f s}(a,b)$ is the contribution of the flavour sensitive scatterings $\nu X\rightarrow \nu X$. This quantity is determined by the square of the difference of the  $\nu_a$-$X$ and $\nu_b$-$X$ scattering amplitudes \cite{Stodolsky:1987dx,Raffelt:1993uj}. In terms of the total scattering rates $ \Gamma(a)$ and $ \Gamma(b)$ one gets \cite{Stodolsky:1987dx}:
\begin{eqnarray} 
\Gamma^{f s}(a,b)\simeq \Gamma(a)+\Gamma(b) -2 \sqrt{ \Gamma(a) \Gamma(b)}~.
\label{eq:lc}
\end{eqnarray}
From eqs. (\ref{eq:15})-(\ref{eq:lc}), using the rates given in ref.\cite{Shi:1993hm} we find:
\begin{eqnarray} 
&&l_c(a, b)=\left[ k(a,b) G^2_F  \alpha^2 \beta^5 T^5 \right]^{-1}~,
\label{eq:16a}
\end{eqnarray}
where, for $T\gta 0.5$ MeV:
\begin{eqnarray} 
&&k(e,\mu)\simeq 6.5 \cdot 10^{-3} \left[16+ 0.5 \left( r(\xi_e)+ r(\xi_\mu)\right)+ 5.3 r(-\xi_e)+3.5 r(-\xi_\mu) \right],
\label{eq:16b}\\
&&k(\mu,\tau)\simeq 6.5 \cdot 10^{-3} \left[0.5 (r(\xi_\mu)+r(\xi_\tau))+ 3.5 (r(-\xi_\mu)+r(-\xi_\tau)) \right]~.
\label{eq:16c}
\end{eqnarray}
Here $r(\xi)\equiv I(\xi)/I(0)$  and $I(\xi)\equiv \int_0^{\infty} x^2 dx/(1+\exp(x-\xi)) $. We define $\xi\equiv \mu/T$ with $\mu$ the chemical potential of the neutrino gas\footnote{Let us recall the relation between the quantity  $\xi$ and the CP-asymmetry $\eta$: $\eta=(\xi^3 + \pi^2 \xi) \beta^3/(12 \zeta(3))$. In eqs. (\ref{eq:16a})-(\ref{eq:16e}) we considered the factor $\alpha$ to have the same value for all the particle species (neutrinos, electrons, positrons). 
We checked that this is a good approximation  even for the large asymmetries we are considering.}. 
At  $T\simeq 0.5$ MeV electrons and positrons annihilate, so that for $T< 0.5$ MeV their contribution to the scattering and absorption rates becomes negligibly small and we get:
\begin{eqnarray} 
&&k(e,\mu)\simeq 6.5 \cdot 10^{-3} \left[ 0.5 \left( r(\xi_e)+ r(\xi_\mu)\right)+ 3( r(-\xi_e)+ r(-\xi_\mu)) \right],
\label{eq:16d}\\
&&k(\mu,\tau)\simeq 6.5 \cdot 10^{-3} \left[0.5 (r(\xi_\mu)+r(\xi_\tau))+ 3 (r(-\xi_\mu)+r(-\xi_\tau)) \right]~.
\label{eq:16e}
\end{eqnarray}

Numerically, eq. (\ref{eq:16a})  gives:
\begin{eqnarray} 
l_c(a, b)\simeq 1.45 \cdot 10^{11}~{\rm cm}~{1 \over k(a,b) \alpha^2 \beta^5} \left({ {\rm 1 MeV}\over T} \right)^5~. 
\label{eq:17}
\end{eqnarray}

\item 
The expansion of the universe.  Oscillations and collisions are ineffective if their scale lengths, $l_v$ and $l_c$, are larger than the inverse expansion rate of the universe, $l_H$.  In the radiation-dominating regime $l_H$ is  expressed as: 
\begin{eqnarray} 
l_H&\simeq& { M_p\over 1.66 \sqrt{g} T^2}~, \nonumber\\
&\simeq & 1.45 \cdot 10^{11}~{\rm cm}{1\over \sqrt{g}}\left({ {\rm 1 MeV}\over T} \right)^2~,
\label{eq:18}
\end{eqnarray}
where  $M_p$ is the Planck mass and $g$ represents the number of relativistic degrees of freedom.  We have $g=10.75$ for $1~ {\rm MeV} \lta T \lta 100~ {\rm MeV}$  and $g=3.36$ for $T\ll 1~ {\rm MeV}$.
\\
\end{itemize}

\noindent

\begin{figure}[hbt]
\begin{center}
\epsfig{file=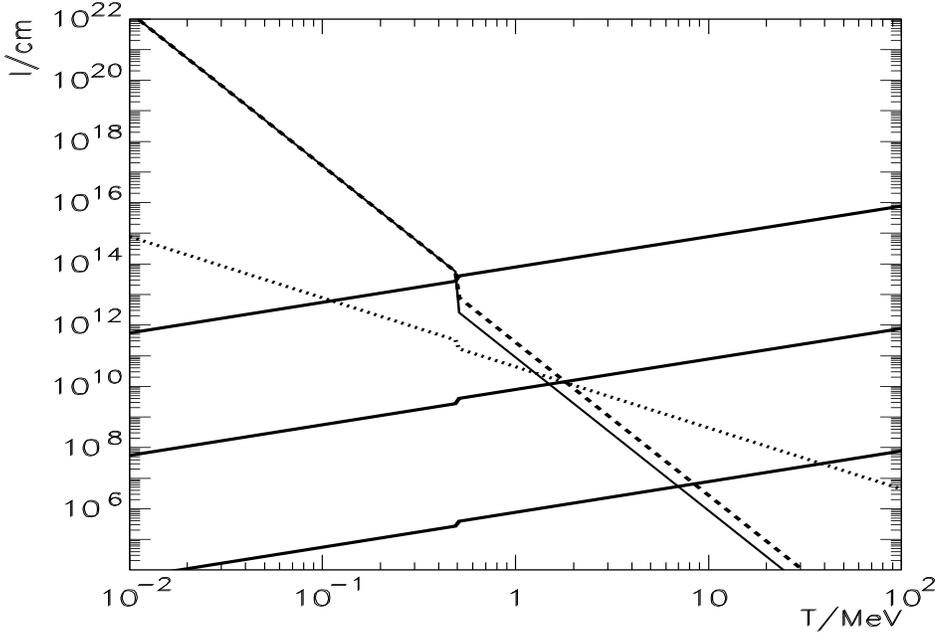, width=14truecm,height=10truecm}
\end{center}
\caption{The  length scales $l_v$, $l_c$  and $l_H$ as functions of the temperature  $T$  of the electromagnetic radiation in the universe. The three thick solid lines represent the vacuum oscillation length $l_v$ and correspond, from the upper to the lower, to $\Delta m^2=10^{-11},10^{-7},10^{-3}~{\rm eV^2}$ respectively. The narrow solid line   and the dashed line represent the coherence length, $l_c$, for the $\nu_e-\nu_\mu$ and  the $\nu_\mu-\nu_\tau$ channels respectively. We have taken $\eta_\mu\simeq 1$, $\eta_e \simeq \eta_\tau\simeq 0$. The dotted line represents the inverse expansion rate of the universe, $l_H$.} 
\label{fig:fig10new} 
\end{figure}

The figure \ref{fig:fig10new} shows the lengths $l_v$, $l_c$  and $l_H$ as functions of the temperature $T$ for $\eta_\mu\simeq 1$, $\eta_e \simeq \eta_\tau\simeq 0$ and various values of $\Delta m^2$.

According to the fig. \ref{fig:fig10new} for $\Delta m^2\simeq 10^{-7}~{\rm eV^2}$
we have $l_v\sim l_c\sim  l_H$ at $T\sim 2~ {\rm MeV}$.  Before this epoch $l_c\lta l_H \lta l_v$, so that collisions are much more efficient than oscillations. 
The inequality $l_c \ll l_v$ implies that  the vacuum oscillation probability, $P_c$, is suppressed by collisions, as we discussed in this section.  As a consequence, for small mixings, $\sin^2 2\theta\ll 1$, the flavour composition of the background remains unchanged until $T\sim 2~ {\rm MeV}$; a partial equilibration of the flavours can be realized for large  mixings: $\sin^2 2\theta\gta 0.5$.
For $T< 2~ {\rm MeV}$ collisions are ineffective, since $l_c\gta l_H$, and vacuum oscillations develop.

With the decrease of $\Delta m^2$, $\Delta m^2\ll 10^{-7}~{\rm eV^2}$, the oscillation length $l_v$ increases and, in consequence, for $T\gta 2~ {\rm MeV}$  the suppression of oscillations due to collisions  is stronger. 
  Even for large mixings  the flavour densities are preserved  until the neutrino decoupling, 
 $T\sim 2~ {\rm MeV}$.  After this epoch oscillations are still suppressed by the expansion rate of the universe and become 
 effective, thus changing the flavour composition of the background, only  when the oscillation length is smaller than the horizon, $l_v\lta l_H$. 

For $\Delta m^2\gta 10^{-7}~{\rm eV^2}$ the inequality  $l_v\lta l_c\lta l_H$ is realized before the decoupling.  In  this circumstance the conversion probability, $P_c$, is not suppressed, in contrast with the case $\Delta m^2\gta 10^{-7}~{\rm eV^2}$.  Collisions are effective, thus leading to the equilibration of the flavour densities even for small mixing angles.  Taking, for instance, $\Delta m^2\simeq 10^{-3}~{\rm eV^2}$ we find that equilibration can be achieved for $\sin^2 2\theta\gta 10^{-3}$. 
Again, after neutrino decoupling the  conversion is determined by vacuum oscillations.
\\

\noindent
For a different choice of the asymmetries at production, e.g. $\eta_\mu \simeq \eta_\tau\simeq 1$ and $\eta_e \simeq 0$, the results are similar to those in fig. \ref{fig:fig10new} and we come to analogous conclusions.
\\

Let us now find the present flavour asymmetries\footnote{In the following sections we will consider relatively recent epochs ($z\lta 50$) for which the flavour composition of the background is well approximated by the present one  (see sections \ref{sec:new} and \ref{sec:3}).},  $\eta^0_e$, $\eta^0_\mu$, $\eta^0_\tau$, for specific neutrino mixings and mass spectra motivated by the oscillation interpretation of the solar and atmospheric neutrino anomalies.

We consider the mixing matrix:
\begin{eqnarray} 
U=\pmatrix{ {c_\theta}  & -{s_\theta}  & 0  \cr
{s_\theta c_\Theta}   &  {c_\theta c _\Theta}     & -{s_\Theta}  \cr
{s_\theta s_\Theta}   &  {c_\theta s_\Theta}      &  {c_\Theta}  \cr}~, 
\label{eq:10}
\end{eqnarray}
where $c_\theta\equiv \cos \theta$  and  $s_\theta\equiv \sin \theta$ and analogous definitions hold for $s_\Theta$ and $c_\Theta$.
The mass eigenstates $\nu_1$,  $\nu_2$ and $\nu_3$ are related to the flavour ones by  the rotation: $\nu_\alpha=\sum_i U_{\alpha,i}\nu_i$.
The mass squared differences $\Delta m^2_{j i}\equiv m^2_j -m^2_i$ are  taken to be $\Delta m^2_{3 2}=\Delta m^2_{atm}\sim 10^{-3}~{\rm eV^2}$ and $\Delta m^2_{2 1}=\Delta m^2_{\odot}\lta 10^{-5}~{\rm eV^2}$ according to the currently favoured solutions of the solar neutrino problem.   
Let us first consider $\Delta m^2_{2 1}< 10^{-7}~{\rm eV^2}$, as predicted by the LOW and VO solutions.  This range of $\Delta m^2_{2 1}$ is the most relevant to the conversion of ultra-energetic neutrinos (see sect. \ref{subsec:4.3}). 

According to the results of this section, we identify the following scenario:
\\

\noindent 
Suppose that before the neutrino decoupling epoch, at $T> 2~ {\rm MeV}$, a large asymmetry has been produced in one flavour while the other asymmetries are initially small:  e.g. $\eta_\mu=2 \eta$ and $\eta_e\simeq \eta_\tau\simeq 0$, with  $\eta\sim 1$. 
As the universe evolves down to $T\sim 2~ {\rm MeV}$,  the muon and tau asymmetries will be equilibrated by the combined effect of oscillations  and collisions. In the same epochs $\nu_e$ oscillations are still suppressed by collisions. Therefore,   
the electron neutrino asymmetry, $\eta_e$, remains unchanged and, 
 at $T\simeq T_{BBN}$, we have $\eta_\mu\simeq\eta_\tau=\eta$  and $\eta_e \simeq 0$.   After this epoch neutrinos decouple from the thermal bath; collisions become ineffective and the system evolves according to vacuum oscillations. During the evolution decoherence occurs due to the spread of the wavepackets.  Therefore at the present epoch the background neutrinos are in mass eigenstates.   With the mixing (\ref{eq:10}) we find the present asymmetries $\eta^0_i$ for these states:
\begin{eqnarray} 
\eta^0_1\simeq\eta  \sin^2 \theta~ ,~~~~~~~~
\eta^0_2\simeq\eta  \cos^2\theta~, ~~~~~~~
 \eta^0_3\simeq\eta ~.
\label{eq:etai}
\end{eqnarray} 
The corresponding flavour asymmetries equal:
\begin{eqnarray} 
&&\eta^0_e\simeq\eta {1\over 2} \sin^2 2\theta~ ,\nonumber\\
&&\eta^0_\mu\simeq\eta  \left(1-{1\over 2}\sin^2 2\theta \cos^2\Theta\right)~ ,\nonumber\\
&&\eta^0_\tau\simeq\eta \left(1-{1\over 2}\sin^2 2\theta \sin^2\Theta\right).
\label{eq:19}
\end{eqnarray} 
One can see that a large asymmetry  is produced in the electron flavour provided that the mixing of the electron neutrino is large: $\sin^2 2\theta \sim 1$.
Thus, the electron neutrino asymmetry at the present epoch  can be much larger than the upper bound  (\ref{eq:bbnbound}).  Notice also that  $\eta^0_\mu\simeq \eta^0_\tau$ for $\Theta\simeq \pi/4$.
\\

\noindent
If equally large asymmetries are initially produced  in the muon and tau flavours, $\eta_\mu=\eta_\tau=\eta\sim 1$ and $\eta_e \simeq 0$,
the equality $\eta_\mu=\eta_\tau$ is preserved until the decoupling epoch, $T\sim 2~ {\rm MeV}$, due to the combined effect of  oscillations and collisions.
The evolution of $\eta_e$ is blocked by collisions for $T\gta 2~ {\rm MeV}$.
After the neutrino decoupling vacuum oscillations take place; with the mixing matrix (\ref{eq:10}) we get  the same results as in eqs.  (\ref{eq:etai})-(\ref{eq:19}).
\\

For $\Delta m^2_{2 1}\simeq 10^{-6}~{\rm eV^2}$ and small mixing, $\sin^2 2\theta \simeq 10^{-4}-10^{-3}$, according to the SMA  solution of the solar neutrino problem, the equilibration length $l_{eq}$ is very large. Therefore  the electron asymmetry $\eta_e$ is not equilibrated with the muon and tau asymmetries. Again, the present asymmetries are determined by vacuum oscillations which occur after the neutrino decoupling and lead to the result (\ref{eq:19}).

Conversely, for large mixing angle, $\sin^2 2\theta \sim 1$ and $\Delta m^2_{2 1}\simeq 10^{-7}-10^{-5}~{\rm eV^2}$, as given by part of the LOW solution and by the LMA solution regions, equilibration is rapidly realized and one gets:
 \begin{eqnarray} 
\eta^0_e\simeq\eta^0_\mu\simeq\eta^0_\tau \simeq {2\over 3}\eta ~.
\label{eq:equil}
\end{eqnarray} 
Notice, however, that the results (\ref{eq:etai})-(\ref{eq:equil}) depend on the epoch we considered for the production of the large CP-asymmetry $\eta$: the equilibration effect of collisions does not take place if the neutrino asymmetries are generated at epochs close to the neutrino decoupling epoch, at $T\sim 1~{\rm MeV}$.

\subsubsection{Evolution in presence of a sterile state}
\label{subsubsec:2.2.2}
If a sterile state, $\nu_s$, is mixed with the three active ones, a general description of the evolution of the neutrino gas is complicated and would deserve a detailed study.

We consider here the specific case in which the sterile neutrino is mixed mainly with one active state only, e.g. $\nu_e$, and the admixture of $\nu_s$ with $\nu_\mu$ and $\nu_\tau$ is negligible.  In other words, we consider the mass states  $\nu_0\simeq \cos \theta \nu_e + \sin \theta \nu_s$ and the orthogonal combination  $\nu_1\simeq -\sin \theta \nu_e + \cos \theta \nu_s$. Similarly, we take  $\nu_2\simeq \cos \varphi \nu_\mu + \sin \varphi \nu_\tau$ and   $\nu_3\simeq -\sin \varphi \nu_\mu + \cos \varphi \nu_\tau$.
This would correspond to  $\nu_e-\nu_s$ solution of the solar neutrino problem 
 and $\nu_\mu-\nu_\tau$ solution of the atmospheric neutrino anomaly.\\

Let us consider the evolution of the  $\nu_e-\nu_s$ system. 
 The effective mixing angle in matter, $\theta_m$, can be written as: 
\begin{eqnarray} 
\tan 2\theta_m={\sin 2\theta\over \cos 2\theta -2EV/\Delta m^2 }~, 
\label{eq:tant}
\end{eqnarray}
where $V$ is given in eq. (\ref{eq:potefunivb}) and 
 $E$ is the average thermal energy of the neutrinos:
 $E\simeq \alpha \beta T$. 
Numerically, from eqs. (\ref{eq:potefunivb}) and (\ref{eq:tant}) we get:
\begin{eqnarray} 
\tan 2\theta_m={\sin 2\theta\over \cos 2\theta -
 0.8\cdot 10^{4} \alpha \beta F \eta_\nu
\left( {T/ 1~{\rm MeV}}\right)^4
\left({10^{-3}~{\rm eV^2}/ \Delta m^2}\right) }~,
\label{eq:tantnum}
\end{eqnarray}
where we used the expression $n_\gamma(T)=2\zeta(3)T^3/\pi^2$ for the concentration of photons at the temperature $T$, with the value $\zeta(3)\simeq 1.202$ for the Riemann zeta function.

From eq. (\ref{eq:tantnum}) it follows that, for $\eta_{\nu}\gta 1$, $T\gta T_{BBN}$ and 
$\Delta m^2\lta 1~{\rm eV^2}$ the mixing is strongly suppressed, 
$\tan 2\theta_m\ll 1$, corresponding to $\theta \simeq \pi/2$ ($\theta \simeq 0$) if $F \eta_\nu>0$ ($F \eta_\nu<0$).
Thus, no level crossing is realized before the BBN epoch. At $T\gta 2~{\rm MeV}$ collisions are effective (see fig. \ref{fig:fig10new}); however they do not modify $\eta_e$ significantly due to the very small value of the mixing and consequently of the conversion probability $P_c$ (see section \ref{subsubsec:2.2.1}, eq. (\ref{eq:damp})).   
Thus, we conclude that no significant flavour conversion occurs and  the original value of $\eta_e$ is  preserved at least until the BBN epoch, even in the case of large vacuum mixing angles.

As the temperature decreases, $T< T_{BBN}$, the mixing angle $\theta_m$ approaches rapidly its vacuum value.  Taking  $\eta_\nu\simeq 1$, $F=2$ and $\Delta m^2\simeq 10^{-3}~{\rm eV^2}$ we get $\tan 2\theta_m-\tan 2\theta \lta 10^{-2}$ for $T\lta 10$ KeV.

Considering that the propagation of the neutrino states is adiabatic (see section \ref{subsec:3.2}), we find the present concentrations of $\nu_e$ and $\nu_s$ in terms of the initial density $n_e$\footnote{We assume that only active  states are initially produced before the BBN epoch, thus $n_s=0$.}: 
\begin{eqnarray} 
&&n^0_s=n_e \cos^2 \theta ~,~~~~~~~~~n^0_e=n_e \sin^2 \theta ~,~~~~~~~~~{\rm if}~ F \eta_\nu>0 
\label{eq:presdens1}\\
&&n^0_s=n_e \sin^2 \theta ~,~~~~~~~~~n^0_e=n_e \cos^2 \theta ~,~~~~~~~~~{\rm if}~ F \eta_\nu<0.
\label{eq:presdens2}
\end{eqnarray}
If  $n_{e}\gg n_{\bar{e}}$, the concentrations of $\bar{\nu}_e$ can be neglected and   relations analogous to (\ref{eq:presdens1})-(\ref{eq:presdens2}) hold for the CP-asymmetries $\eta^0_{e}$ and  $\eta_{e}$.

The present $\nu_\mu$ and $\nu_\tau$ asymmetries can be found according to
the  discussion in section \ref{subsubsec:2.2.1}. The effect of collisions leads to equilibration of $\eta_\mu$ and $\eta_\tau$ at $T\simeq T_{BBN}$: $\eta_\mu\simeq \eta_\tau=\eta$. At later epochs vacuum oscillations develop, leaving this equality unchanged.
Thus, we can summarize the present  CP-symmetries for the four flavours as follows:
\begin{eqnarray} 
&&\eta^0_s=\eta_e \cos^2 \theta ~,~~~~~~~~~\eta^0_e=\eta_e \sin^2 \theta ~,~~~~~~~~~{\rm if}~ F \eta_\nu>0~, 
\label{eq:presas1}\\
&&\eta^0_s=\eta_e \sin^2 \theta ~,~~~~~~~~~\eta^0_e=\eta_e \cos^2 \theta ~,~~~~~~~~~{\rm if}~ F \eta_\nu<0~,
\label{eq:presas2}\\
&&\eta^0_\mu\simeq \eta^0_\tau\simeq \eta~.
\label{eq:presas}
\end{eqnarray}
Having neglected any mixing between $\nu_e$ and the other (active) flavours, we find that the present value of the electron neutrino asymmetry is smaller than the one at the BBN epoch, $\eta^0_e\leq \eta_e$, thus remaining within the bound given in (\ref{eq:bbnbound}).


\section{High energy neutrino conversion: the active-active case}
\label{sec:new}

Let us consider three mixed active neutrinos, $\nu_e$, $\nu_\mu$, $\nu_\tau$, and find the  potential for a beam of high energy neutrinos (``beam neutrinos'') due to the interaction with the relic neutrino background (``background neutrinos'').
As discussed in the section \ref{sec:2}, the flavour composition of the neutrino background changes with time due to the neutrino mixing. However,  we will focus on  neutrinos produced in relatively recent epochs, $z\lta 50$, when 
 the flavour content of the 
relic neutrino gas has already settled down and does not change with time.


\subsection{The refraction potential}
\label{subsec:new.1}
According to the section \ref{sec:2}, decoherence due to the spread of wavepackets implies that the background neutrinos are in mass eigenstates.    
As a consequence, the matrix  of potentials, $V_\nu$, for the beam neutrinos propagating in this background is not diagonal in the flavour basis, $(\nu_e,\nu_\mu,\nu_\tau)$. It is possible to check \cite{Pantaleone:1992eq} that $V_\nu$ becomes diagonal in the basis of the mass eigenstates, $( \nu_1,\nu_2,\nu_3 )$, where it can be written as:
\begin{eqnarray} 
V_\nu&=&\sqrt{2} G_F \left[(n_1-n_{\bar 1})+(n_2 -n_{\bar 2}) + (n_3 -n_{\bar 3})\right] 
\label{eq:potmass}\\
&+& \sqrt{2} G_F \pmatrix{ {n_1 f(-s^{(1)}_Z)-n_{\bar 1} f(s^{(1)}_Z)}  & 0  & 0  \cr
0   &    n_2 f(-s^{(2)}_Z)-n_{\bar 2}f(s^{(2)}_Z)    & 0 \cr
0   &   0      &  n_3 f(-s^{(3)}_Z)-n_{\bar 3}f(s^{(3)}_Z) \cr}~,\nonumber
\end{eqnarray}
where $n_i$ ($n_{\bar i}$) denotes the concentration of the mass state $\nu_i$ ($\bar{\nu}_i$) in the background and
 $f(s^{(i)}_Z)$ is the $Z$-boson propagator function:
\begin{equation}
f(s^{(i)}_Z)\equiv{1-s^{(i)}_Z \over (1-s^{(i)}_Z)^2+\gamma^2_Z }~. 
\label{eq:ffunc} 
\end{equation}
Here  $\gamma_Z$ and $s^{(i)}_Z$ are the
normalized width of the $Z$-boson and  total energy squared in the $\nu_i-\nu_i$ center 
of mass for non-relativistic background neutrinos:
\begin{eqnarray}
s^{(i)}_Z\simeq {2E m_i \over M^2_Z}\simeq 2.4\cdot 10^{-2} 
\left({E \over 10^{20}{\rm eV}}\right )
\left({m_i \over 1~{\rm eV}}\right )~,~~~~~\gamma_Z\equiv {\Gamma_Z \over
M_Z}~,
\label{eq:param} 
\end{eqnarray}
with  $m_i$ being the mass of the neutrino $\nu_i$ and $E$ the energy of the beam neutrino.

The terms in the first line of eq. (\ref{eq:potmass})  
are due to neutral current $\nu-\nu$ scattering in the $t$-channel.
The terms in the second line of eq. (\ref{eq:potmass}) represent the contributions of  $\nu_i-\nu_i$   scattering with $Z$-boson exchange 
in the $u$-channel, and of $\nu_i-{\bar \nu}_i$ annihilation processes.

For $E\lta 10^{20}{\rm eV}$ and $m_i\lta 1~{\rm eV}$ the energy in the
$\nu_i-\nu_i$ center of mass is much below the $Z$-boson resonance: $s_Z\lta 0.03$.
In this case the propagator function (\ref{eq:ffunc}) reduces to unity:
$f(s^{(i)}_Z)\simeq f(-s^{(i)}_Z)\simeq 1$, and the neutrino-neutrino potential (\ref{eq:potmass}) becomes energy-independent.

For extremely high energies,
$E\simeq 10^{21}-10^{22} {\rm eV}$, and neutrino mass of order
$1~{\rm eV}$  the propagator corrections become important.  However, in this
range of energies the absorption effects of the neutrino background are
strong \cite{Lunardini:2000sw,Roulet:1993pz}. Therefore, the neutrino fluxes at Earth 
are largely suppressed.  
In what follows we will concentrate on the low energy limit, $s^{(i)}_Z\ll 1$, which is mainly relevant for applications.

For a beam of antineutrinos propagating in a neutrino background the potential
$V_{\bar {\nu}}$ is given by eq. (\ref{eq:potmass}) with the replacement
$n_{i}\rightarrow n_{\bar{i}}$ and vice-versa for all the $\nu_i$ states.
\\

The fact that the neutrino-neutrino potential matrix, eq. (\ref{eq:potmass}), is diagonal in the basis of mass eigenstates has a straightforward consequence: the effect of refraction consists in a modification of the neutrino effective masses only.  In terms of the present  CP-asymmetries $\eta^0_i$ for the mass states $\nu_i$  of the background  
we find (for $s^{(i)}_Z\ll 1$) the following corrections:
\begin{eqnarray} 
&&{\Delta m_{2 1}^2 \over 2 E}\rightarrow {\cal E}_{2 1}\equiv { \Delta m_{2 1}^2 \over 2 E}  
 + \sqrt{2} G_F n_\gamma (\eta^0_2-\eta^0_1)~,\nonumber \\
&&{\Delta m_{3 2}^2 \over 2 E}\rightarrow {\cal E}_{3 2}\equiv { \Delta m_{3 2}^2 \over 2 E}  
 + \sqrt{2} G_F n_\gamma (\eta^0_3-\eta^0_2)~.
\label{eq:9}
\end{eqnarray}

As we discussed in section \ref{sec:2}, the present composition of the neutrino background is determined by the initial flavour asymmetries, $\eta_e$, $\eta_\mu$ and $\eta_\tau$, and by the mixing matrix $U$ of the neutrino system.  
The expressions of ${\cal E}_{2 1}$ and  ${\cal E}_{3 2}$ in terms of these quantities can be found from the results of section  \ref{subsubsec:2.2.2}.  
In particular, with the asymmetries (\ref{eq:etai}) eq. (\ref{eq:9}) gives:
\begin{eqnarray} 
&&{\cal E}_{j i}= {\Delta m_{j i}^2 \over 2 E} + V_{j i}, 
\label{eq:epssynt}\\
&&V_{j i}\equiv F_{j i} \eta \sqrt{2} G_F n_\gamma  ~, 
\label{eq:vdef}\\
&&F_{2 1}=\cos 2\theta~, ~~~~~~~~~F_{3 2}=\sin^2 \theta~.
\label{eq:fji}
\end{eqnarray}
Here we denote as $\eta$ the maximal flavour asymmetry, $\eta\equiv Max\{\eta_{\mu},
\eta_{\tau},\eta_{e} \}$, which is realized in the background at the epoch of nucleosynthesis, $T\simeq T_{BBN}$; thus $\eta$ is constrained by the bounds (\ref{eq:bbnbound}).

In eqs. (\ref{eq:epssynt})-(\ref{eq:vdef}) 
the information on the specific mixing matrix and initial composition of the background are encoded in the $F$ factors. The dependence of $F_{j i}$  on the mixing angle in eq. (\ref{eq:fji}) is a consequence of expressing the potential (\ref{eq:vdef}) is terms of the flavour asymmetry $\eta$, while the background neutrinos are in mass eigenstates.
For simplicity in what follows we will drop the indexes $j,i$ from the quantities ${\cal E}$, $V$ and $F$ in the expressions  (\ref{eq:epssynt})-(\ref{eq:fji}).


\subsection{The conversion probability}
\label{subsec:new.2}
From the fact that the potential $V$ modifies  the effective mass eigenvalues, eqs. (\ref{eq:epssynt})-(\ref{eq:fji}), it follows that the interaction with the neutrino background does not change the mixing matrix of the neutrino system, which remains the same as in vacuum.  Conversely, the phase of oscillations is affected by the medium, so that  the dynamics of the neutrino propagation consists in oscillations with constant depth, given by the vacuum mixing angle, and varying oscillation length. 
The  probability $P$ of conversion between two active neutrinos, $\nu_\alpha$, $\nu_\beta$, with mixing angle $\theta$ equals:
\begin{eqnarray} 
P(t,t_i)=\sin^2 2\theta \sin^2 \left( {\Phi \over 2} \right)~,
\label{eq:vosc}
\end{eqnarray}
and the oscillation phase $\Phi$ is given by:
\begin{eqnarray} 
\Phi(t,t_i)=\int_{t_i}^{t} {\cal E}(\tau) d\tau~.
\label{eq:eq2}
\end{eqnarray}
We denote as $t_i$, $t$  the initial and final time of the evolution of the system;  ${\cal E}$ is given in eq. (\ref{eq:epssynt}).  
Using the scaling relations: 
\begin{eqnarray} 
 E=E_0(t_0/t)^{2/3}=E_0(1+z)~, \hskip 0.3cm V=V_0 (t_0/t)^2= V_0 (1+z)^3~, 
\label{eq:scale} 
\end{eqnarray}
where $E_0$ and $V_0$ are the energy and the potential at the present epoch, $z= 0$, we get:
\begin{eqnarray} 
\Phi=\Phi_{vac}+\Phi_{matt}~,
\label{eq:eq3}
\end{eqnarray}
with the following expressions for the vacuum oscillation phase, $\Phi_{vac}$, and  the matter contribution $\Phi_{matt}$:
\begin{eqnarray} 
&&\Phi_{vac}(x,x_i)={3 \over 10}{\Delta m^2 t_0 \over E_0 } \left(x^{ 5\over 3} -x_i^{ 5\over 3}\right)~,
\label{eq:phiv}\\
&&\Phi_{matt}(x,x_i)= V_0 t_0 \left({1\over x_i} - {1\over x} \right)~.
\label{eq:deltaphi}
\end{eqnarray}
We defined $x\equiv t/t_0$  and $x_i\equiv t_i/t_0$.

The matter induced phase, $\Phi_{matt}$, depends only  on the characteristics of the background and on the initial and final moments of time.  
In particular for early production epochs, $x_i\ll 1$, one gets:
\begin{eqnarray} 
\Phi_{matt} \simeq  V_0 t_0 {1\over x_i} ~,
\label{eq:eq5}
\end{eqnarray}
which shows that the phase $\Phi_{matt}$  is accumulated mainly at the production time.

Being independent of $E_0/\Delta m^2$, the  phase $\Phi_{matt}$ becomes comparable or even larger than the  vacuum oscillation phase, $\Phi_{vac}$, at very high energies, $E_0/\Delta m^2 \gta 10^{32}~{\rm eV^{-1}}$.
Taking $x_i=0.125$, corresponding to production at redshift $z\simeq 3$, and $x=1$ we have $\Phi_{vac}\simeq 0.23 \pi$ for  $E_0/\Delta m^2 \simeq 10^{32}~{\rm eV^{-1}}$. This is comparable to the matter phase, $\Phi_{matt} \simeq 0.29 \pi$ given by eqs. (\ref{eq:deltaphi}) and (\ref{eq:vdef}) with $F\eta\simeq 10$.   As $E_0/\Delta m^2$ increases the vacuum phase $\Phi_{vac}$ decreases and the total oscillation phase is dominated by the matter contribution $\Phi_{matt}$.
From eqs. (\ref{eq:vosc}) and (\ref{eq:deltaphi}) we find the asymptotic value of the conversion probability\footnote{
Even though eq. (\ref{eq:asympprob}) gives a a non-zero value for the conversion probability in the limit $\Delta m^2 \rightarrow 0$, the matter effect we are describing requires massive non-degenerate neutrinos: $\Delta m^2 \neq 0$. This condition is necessary for our starting point (section \ref{subsec:new.1}, see eq. (\ref{eq:potmass})) that the neutrinos in the background are in mass eigenstates (different from the flavour ones), produced from flavour states by the spread of the wavepackets during the evolution of the universe.  Thus, the expression (\ref{eq:asympprob}) should be intended as the high-energy limit of the conversion probability for a given (non-zero) value of $\Delta m^2$. 
}:
\begin{eqnarray} 
P(E_0/\Delta m^2 \rightarrow \infty) =\sin^2 2\theta \sin^2 \left(  {1 \over 2 } V_0 t_0 \left({1\over x_i} - {1\over x} \right) \right)~. 
\label{eq:asympprob}
\end{eqnarray}
Notice that the expression (\ref{eq:asympprob}) is insensitive to the change of sign of the potential $V_0$ (i.e. of the product $F\eta$): this implies that in the limit of very high energies a beam of neutrinos and one of antineutrinos will experience the same matter effect. 

\begin{figure}[hbt]
\begin{center}
\epsfig{file=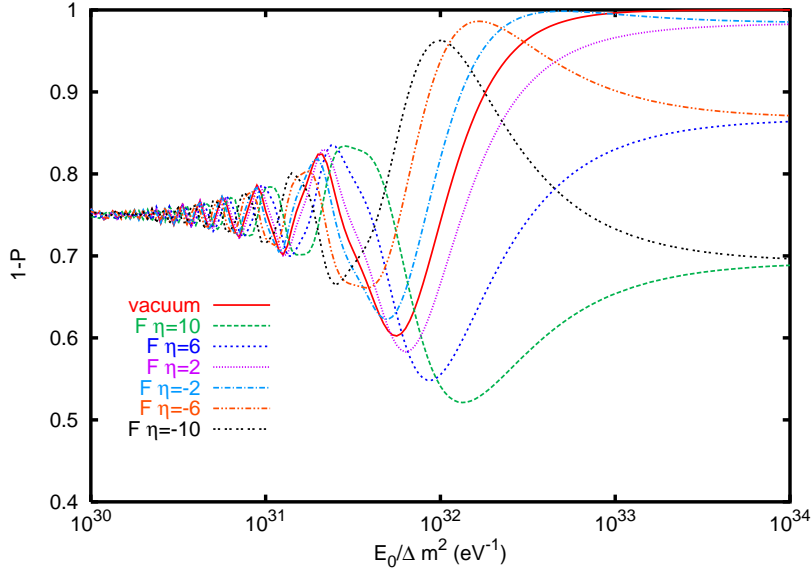, width=11truecm}
\end{center}
\caption{The survival probability $1-P(\nu_\alpha\rightarrow \nu_\beta)$ as a function of the ratio $E_0/\Delta m^2 $ for various values of  $F \eta$. 
We have taken $\sin^2 2\theta=0.5$ and production epoch $z=3$. } 
\label{fig:fig8new} 
\end{figure}

In fig. \ref{fig:fig8new} we show the survival probability, $1-P$, as a function of $E_0/\Delta m^2$ for neutrinos produced at $z=3$ and arriving at Earth at the present epoch, with $\sin^2 2\theta=0.5$ and various values of the product $F\eta$.
The figure was produced by averaging the conversion probability, eq. (\ref{eq:vosc}), over the interval $\Delta E_0\simeq E_0$, keeping in mind the finite accuracy in the reconstruction of the neutrino energy in the detector: 
\begin{equation}
P(E_0)={1\over \Delta E_0}\int_{E_0/2}^{3E_0/2} dE' P(E')~.
\label{eq:aver}
\end{equation} 
  
Let us comment on  the figure \ref{fig:fig8new}.  In absence of asymmetry, $F\eta=0$, the conversion is given by vacuum oscillations ($\Phi_{matt}=0$). For $E_0/\Delta m^2 \gta 10^{32}~{\rm eV^{-1}}$ the vacuum oscillation phase $\Phi_{vac}$ is  very small, thus the conversion probability approaches the unity.  

For strongly CP-asymmetric neutrino background the matter induced phase $\Phi_{matt}$ is sizeable. In consequence, the deviation of the survival probability $1-P$ from the value given by vacuum oscillations can be as large as $\sim 30\%$.  Clearly, as it follows from eq. (\ref{eq:vosc}), a strong effect requires a large mixing angle: if $\sin^2 2\theta\ll 1$ the effect of matter on the oscillation phase will be unobservable due to the very small amplitude of oscillations.
At extremely high energies, $E_0/\Delta m^2 \gta 10^{33}~{\rm eV^{-1}}$, the deviation is constant and independent of the sign of $F\eta$, according to eq. (\ref{eq:asympprob}).


\section{High energy neutrino conversion: the active-sterile case}
\label{sec:3}
Let us consider the case in which a sterile neutrino $\nu_s$ is mixed with the active flavours, $\nu_\alpha$.  We discuss here two-neutrino mixing;  a generalization to a four-neutrino framework will be given in section \ref{subsec:4.3}. 


\subsection{The refraction potential}
\label{subsec:3.1}
In contrast to the active-active conversion studied in section \ref{sec:new},
for an active-sterile neutrino system the matrix of the refraction potentials is diagonal in the flavour basis $( \nu_\alpha,\nu_s )$: 
\begin{eqnarray} 
&&V_\nu=\pmatrix{ {V_\alpha}  & 0   \cr
0   &  0  \cr}~. 
\label{eq:potst}
\end{eqnarray}
In the low energy limit, $s_Z\ll 1$, the potential $V_\alpha$ depends on the flavour asymmetries $\eta^0_e$, $\eta^0_\mu$ and $\eta^0_\tau$ as follows:
\begin{eqnarray} 
&&V_{\alpha}= \sqrt{2} G_{F} n_\gamma \left[  
\eta^0_{{\alpha}}
+\sum_{\beta=e,\mu,\tau} \eta^0_\beta \right]~.
\label{eq:potnumu}
\end{eqnarray}
The potential (\ref{eq:potnumu}) can be written in the same general form as eq. (\ref{eq:vdef}):
\begin{eqnarray}
&&V=F \eta\sqrt{2} G_{F}n_{\gamma}~,
\label{eq:potefuniv} 
\end{eqnarray}
with the same definition of  $\eta$ as  the maximal flavour asymmetry in the background at the epoch of nucleosynthesis, $\eta\equiv Max\{\eta_{\mu},
\eta_{\tau},\eta_{e} \}$. The factor $F$ depends on the specific conversion channel and the flavour content of the background:
\begin{eqnarray} 
F\equiv {1 \over \eta}\left[  
\eta^0_{{\alpha}}
+\sum_{\beta=e,\mu,\tau} \eta^0_\beta \right]~,~~~~~~~~~~{\rm for } ~\nu_\alpha-\nu_s~{\rm channel}~.
\label{eq:deff}
\end{eqnarray}
Let us consider for instance the  conversion of $\nu_e$ to $\nu_s$. If the 
$\nu_e-\nu_s$ conversion  in the background occurred in the resonant channel, the present electron neutrino asymmetry is given by eq.  (\ref{eq:presas1}). 
Using this result and eq. (\ref{eq:presas}) for $\eta^0_{{\mu}}$ and $\eta^0_{{\tau}}$  we get: 
\begin{eqnarray} 
F=2 \left(1+ {\eta_e \over \eta} \sin^2 \theta \right)~,
\label{eq:feex}
\end{eqnarray}
where we considered $\eta_\mu,\eta_\tau\geq \eta_e$.
If the 
$\nu_e-\nu_s$ conversion  in the background proceeded in the non-resonant channel  $\eta^0_e$ is given in eq. (\ref{eq:presas2}).  With this expression one finds a  form for the factor $F$ analogous to eq. (\ref{eq:feex}) with the replacement $ \sin^2 \theta \rightarrow \cos^2 \theta$.

For conversion of antineutrinos the
$\bar{\nu}-\nu$ potential has opposite sign:
$V_{\bar{\nu}}=-V_{\nu}$.  Thus the expression (\ref{eq:potefuniv}) 
holds with the replacement  $F\rightarrow -F$.


\subsection{The dynamics of  neutrino conversion}
\label{subsec:3.2}
 Due to the expansion of the universe the cosmological neutrinos experience a potential which changes with time.   In contrast with the active-active case, the effect of medium changes both the oscillation length and the mixing, eq. (\ref{eq:tant}).
The dynamics of the flavour transition  is 
determined by the resonance and adiabaticity conditions.

Consider the resonance condition: 
\begin{eqnarray}
{2E V \over \Delta m^2}{1\over \cos 2\theta}=1~.
\label{eq:rescond}
\end{eqnarray} 
Using eq. (\ref{eq:potefuniv}) and the scaling relations (\ref{eq:scale}), from eq. (\ref{eq:rescond}) we get the following relations:
\begin{itemize}
\item
The present energy of neutrinos which cross the resonance at the epoch $z$ equals:
\begin{eqnarray}
E_0 = 10^{20} {\rm eV}
\left({\Delta m^2\over 10^{-10} {\rm eV^2}}\right)
{{10^4 \cos 2\theta}\over {F \eta (1+z)^{4}} }~.
\label{eq:resnum} 
\end{eqnarray}

\item
For a given $E_0$ and $\Delta m^2$ the redshift $z_R$ at which the
resonance condition was realized is given by:
\begin{eqnarray}
1+z_R= 10\left[ {{ \cos 2\theta}\over {F \eta}
\left({10^{-10} {\rm eV^2}/ \Delta m^2}\right)
\left(E_0 / 10^{20} {\rm eV}\right)}\right]^{1\over 4}~. 
\label{eq:zr} 
\end{eqnarray}

\item
Neutrinos produced at a distance $z$ undergo resonance if their present energy is in the interval:
\begin{eqnarray}
E_0 = 10^{20} {\rm eV}
\left({\Delta m^2\over 10^{-10} {\rm eV^2}}\right)
{{10^4 \cos 2\theta}\over {F \eta} }
\left[
{{1}\over { (1+z)^{4}} } ~,~1\right]~.
\label{eq:erange} 
\end{eqnarray}
Taking $F \eta=10$, $\cos 2\theta \simeq 1$, $\Delta m^2=10^{-10} {\rm eV^2}$ and $z=3$  from eq. (\ref{eq:erange}) we find  $E_0\simeq 4\cdot 10^{20} -
 10^{23} {\rm eV}$.  With the same values of the parameters and 
$E_0=10^{20} {\rm eV}$ we get that the resonance condition (\ref{eq:resnum}) is satisfied at $z_R\simeq 4.6$. 
\\
\end{itemize}

\noindent
The adiabaticity condition involves the time
variation of both the neutrino energy and the concentration of the neutrino background. It can be expressed it terms  of the
adiabaticity parameter at resonance, $\chi_R$, as:
\begin{eqnarray}
&&\chi_R\gg 1 \nonumber \\
&&\chi_R\equiv \left.{(\Delta m^2)^2 \over 4 E }
\sin^2 2 \theta
\left[ {d\over dt}(E V) \right]^{-1}\right|_{res}~, 
\label{eq:adiab} 
\end{eqnarray}
where the  subscript ``$res$'' indicates that the various quantities are evaluated at resonance, i.e. when the  condition (\ref{eq:rescond}) is fulfilled.
With the potential  (\ref{eq:potefuniv}), using  the scalings (\ref{eq:scale}) and the resonance condition (\ref{eq:rescond}), we find:
\begin{eqnarray}
\chi_R\simeq  10^{-2}F \eta \tan^2 2\theta (1+z_R)^{3\over 2}~.
\label{eq:adiabresnum} 
\end{eqnarray}
For $F \eta\lta 10$, $\tan^2 2\theta=1$ 
and $z_R\lta 5$ one finds $\chi_R\lta 1.4$. Thus, for neutrinos produced at
epochs $z<5$, we expect breaking of the adiabaticity. 
Notice that $\chi_R$ does not depend explicitly on the neutrino energy and mass squared
difference; it increases with $\eta$ and 
$z_R$. 

From eq. (\ref{eq:adiabresnum}) we get the redshift $z_a$ corresponding to
$\chi_R=2\pi\gg 1$:
\begin{eqnarray}
1+z_a\equiv\left[{2\pi\cdot 10^2 \over  F \eta \tan^2
2\theta}\right]^{2\over 3}~. 
\label{eq:za} 
\end{eqnarray}
If the resonance condition is fulfilled at $z\geq z_a$ the level crossing
(resonance) proceeds adiabatically. Taking $F \eta=10$
and $\tan 2\theta=1$ we find $z_a\simeq 15$.

For $\eta \gta 1$ and $\tan 2\theta\leq 1$ we have  $z_a\geq z_d$.  
Thus, we can define three epochs of neutrino production, corresponding to different characters of the 
evolution of the neutrino beam:
\\

\noindent 
(i) Earlier epoch: $z>z_a$, when both adiabaticity and the minimum width
conditions are satisfied. If also the resonance condition is fulfilled at
$z_R>z_a$ the neutrinos will undergo strong resonance conversion.  Otherwise, if
the resonance condition is not realized (e.g., due to a large value of $\Delta
m^2/E$) the matter effect can be small.
\\

\noindent 
(ii) Intermediate epoch: $z_a>z>z_d$. The adiabaticity at resonance is not
satisfied (if $z_a>z_R>z_d$). At the same time the matter width can be large
enough to induce significant matter effect. 

Two remarks are in order. (1) The propagation still can be adiabatic in the part of
the interval $[z_d,z_a]$ outside the resonance, and in the whole of it if the
resonance condition is never satisfied in this time interval.  
(2) In monotonously varying density the condition for strong matter effect reduces
to the adiabaticity condition \cite{Lunardini:2000sw}.  Therefore, in spite of
the fulfillment of the minimum width condition, the matter effect can be small
for neutrinos produced in the most part of the interval  $[z_d,z_a]$.
\\

\noindent
(iii) Later epoch: $z<z_d$.  For neutrinos produced in this epoch the matter
effects are expected to be small. 
\\

\begin{figure}[hbt]
\begin{center}
\epsfig{file=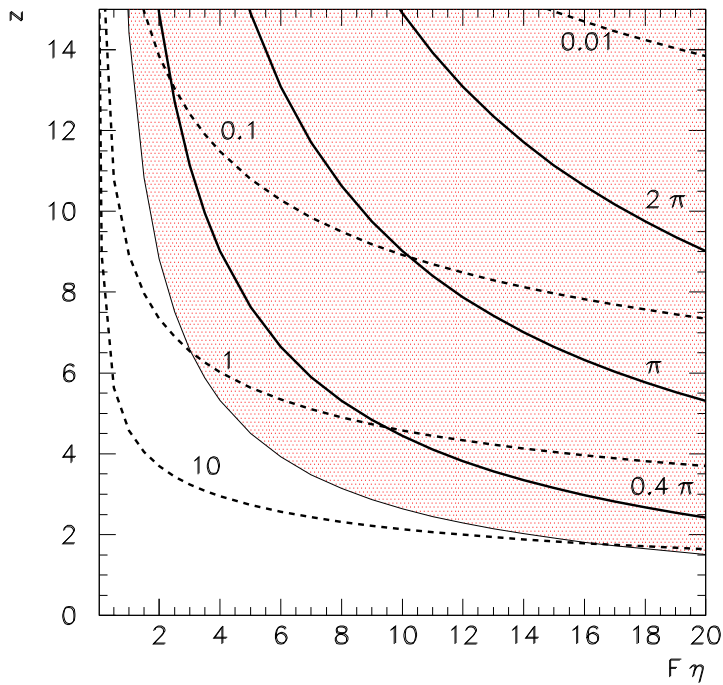, width=11truecm}
\end{center}
\caption{The minimum width, resonance and adiabaticity conditions in the $z$-$F\eta$ plane for $\nu_\alpha-\nu_s$ conversion. The solid lines are iso-contours of adiabaticity, i.e. of the quantity  $\chi_R/\tan^2 2\theta$ (numbers on the curves). The dashed lines are iso-contours of resonance, i.e. of the ratio $E_0/(\Delta m^2 \cos 2\theta)$; the values are given on the curves in units of $10^{30}~{\rm eV^{-1}}$. The minimum width condition is satisfied in the shadowed region. 
} 
\label{fig:fig1} 
\end{figure}

The fig. \ref{fig:fig1} shows the minimum width, resonance and adiabaticity conditions in the $z$ - $F \eta$ plane.  
The minimum width condition (\ref{eq:dasym}) is fulfilled in the shadowed region. The lower border of this area corresponds to the curve $z=z_d(F \eta)$ (eq. (\ref{eq:zd})).
For values of $F \eta$ and of $z$ in this region one may expect significant matter effect.

The dashed lines show the values of $z$ and $F \eta$ for which the resonance condition (\ref{eq:rescond}) is satisfied for neutrinos with a given  $E_0/(\Delta m^2 \cos 2\theta)$ (iso-contours of resonance).

The solid lines are iso-contours of the adiabaticity parameter: they are contours of constant ratio $\chi_R/\tan^2 2\theta$ (see eq. (\ref{eq:adiabresnum})). The upper curve corresponds to  $\chi_R/\tan^2 2\theta=2\pi$, that is, to $z=z_a$ for $\tan^2 2\theta=1$. For values of neutrino production epoch $z$ and $F \eta$ above this contour one would expect resonant adiabatic conversion as dominating mechanism of neutrino transformation.   For a given $F \eta$ and $z$ the adiabaticity iso-contour gives the value of $\chi_R/\tan^2 2\theta$ for neutrinos produced at the epoch $z_i\geq z$ and having the resonance at $z$. 
In turn, the resonance at $z$ and $F \eta$ can be satisfied for certain values of   $E_0/(\Delta m^2 \cos 2\theta)$.
It is clear from the figure that strong adiabatic conversion occurs for large production epochs, $z\gta 10$, large asymmetry, $F \eta\gta 10$ and large mixing $\tan^2 2\theta\sim O(1)$.  For $F \eta\sim 2$ the minimal width condition is fulfilled for large production epochs, $z\gta 8$, and some effects of adiabatic conversion may be seen at  $z\gta 15$.

From the above considerations it appears that for realistic parameters a
flavour transition of neutrinos occurs either due to vacuum oscillations
modified by matter effect or by an interplay of oscillations and non-adiabatic
conversion.


\subsection{The conversion probability}
\label{subsec:3.3}

Let us  consider neutrinos produced at a given epoch $z$ with a certain flavour $\nu_\alpha$ and
propagating in the expanding universe with a given constant asymmetry
$\eta$. 

We find the $\nu_\alpha-\nu_s$ conversion probability by numerical solution of the
evolution equation for two neutrino species with the Hamiltonian in the flavour
basis $(\nu_\alpha,\nu_s)$:
\begin{equation}
H=\pmatrix{-{\Delta m^2 \over 2E(z)}\cos 2\theta +V(z)  
& {\Delta m^2 \over 4E(z)}\sin 2\theta\cr 
  {\Delta m^2 \over 4E(z)}\sin 2\theta    & 0 \cr}~,
\label{eq:hamilt}
\end{equation} 
where $E(z)$ and $V(z)$ scale according to eq. (\ref{eq:scale}).

\begin{figure}[hbt]
\begin{center}
\epsfig{file=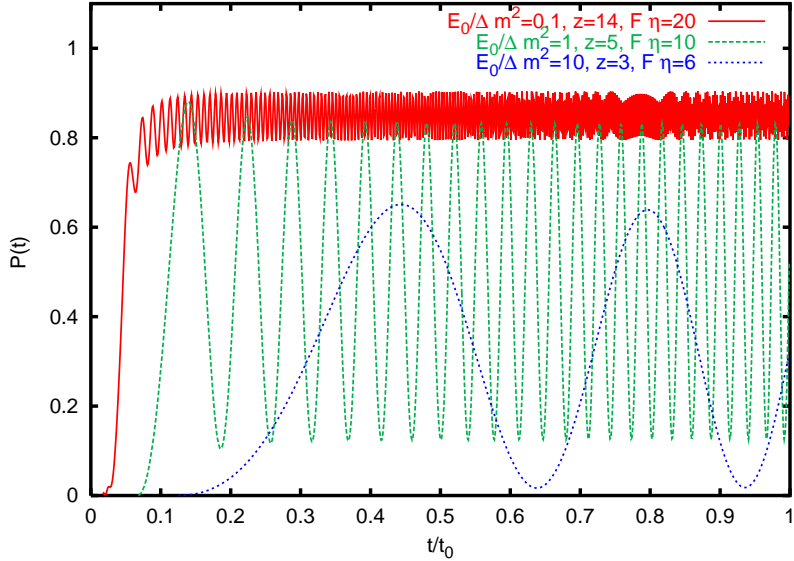, width=11truecm}
\end{center}
\caption{The $\nu_\alpha - \nu_s$  conversion probability $P(t)$ as a function of time. 
We have taken $\sin^2 2\theta=0.5$ and three different choices of $E_0/\Delta m^2 $ (in units of $10^{30}~{\rm eV^{-1}}$), production epoch $z$ and $F \eta$. The time $t$ is given in units 
 of the age of the universe, $t_0$. 
} 
\label{fig:fig2} 
\end{figure}

 As discussed in sect. \ref{subsec:3.2} (see fig. \ref{fig:fig1}), the dynamics of flavour transformation depends on the production epoch $z$, the resonance epoch, $z_R$, which depends on $E_0/\Delta m^2$, and on the value of the adiabaticity parameter at resonance,  $\chi_R$.   The figure \ref{fig:fig2} illustrates the real time evolution of the neutrino states for $\sin^2 2\theta=0.5$ and different $z,z_R,\chi_R$, which represent different regimes of conversion.  

The solid curve corresponds to production much before the resonance epoch: $z>z_R=6.8$ and weak adiabaticity breaking in the resonance, $\chi_R\simeq 4.3$.  The dominating process is the adiabatic conversion which occurs in the resonance epoch, $t_R/t_0\simeq 0.05$.  The averaged transition probability is close to what one would expect for the pure adiabatic case: $P_{ad}=1-\sin^2 \theta=0.85$. Weak adiabaticity violation leads to the appearance of oscillations at $t>t_R$.

The dashed curve corresponds to production close to resonance $z\gta z_R=4.2$ and strong adiabaticity violation in the resonance: $\chi_R\simeq 1.2$. The dominating process is oscillations in matter with resonance density. At production the mixing is almost maximally enhanced, $\sin^2 2\theta_m\simeq 1$. The change of matter density leads to slight increase of the average conversion probability with respect to  $\sin^2 2\theta_m/2$. The decrease of density is fast: the typical scale of density change is smaller than the oscillation length, so that maximal depth oscillations do not have time to develop.   

The dotted line shows the same type of regime with stronger adiabaticity violation in resonance.  The depth $D$ of oscillations is smaller, and the average conversion probability is close to $D/2$.    

\begin{figure}[hbt]
\begin{center}
\epsfig{file=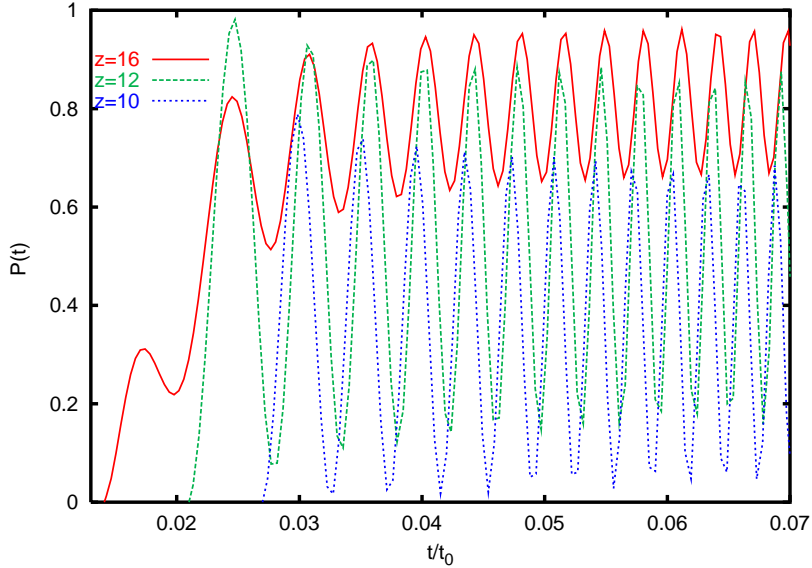, width=11truecm}
\end{center}
\caption{The $\nu_\alpha - \nu_s$ conversion probability $P(t)$ as a function of time in the regime of good adiabaticity (see fig. \ref{fig:fig1}).
We have taken production epochs $z$ earlier, simultaneous and later than the resonance epoch $z_R$. Here
$\sin^2 2\theta=0.5$, $F \eta=14$ and 
$E_0/\Delta m^2 =1.8\cdot 10^{28}~{\rm eV^{-1}}$.   The time $t$ is given in units 
 of the age of the universe, $t_0$. 
} 
\label{fig:fig2a} 
\end{figure}

For further illustration, in fig. \ref{fig:fig2a} we show the evolution in the case of good adiabaticity. Different curves correspond to different production epochs: (i) before resonance, $z>z_R$, (ii) at resonance, $z=z_R$, (iii) after resonance, $z<z_R$.

\begin{figure}[hbt]
\begin{center}
\epsfig{file=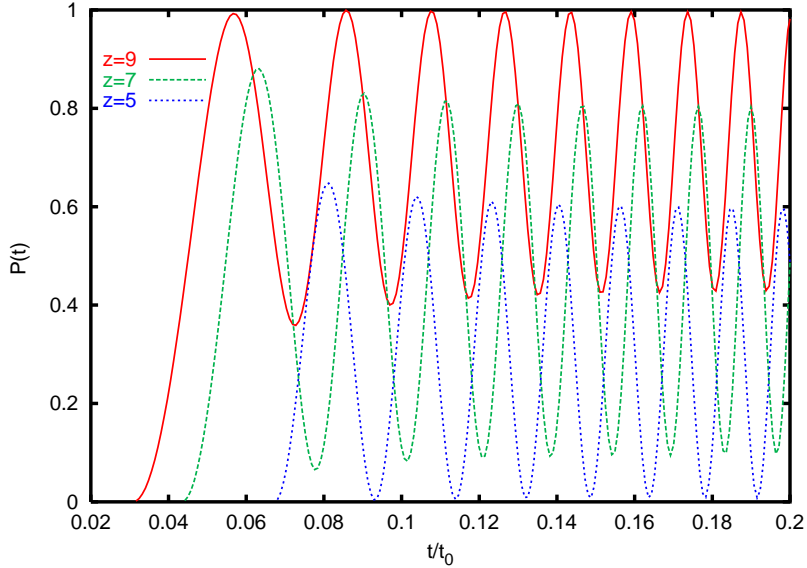, width=11truecm}
\end{center}
\caption{The same as fig. \ref{fig:fig2a} for the regime of moderate breaking of adiabaticity. Here
$\sin^2 2\theta=0.5$, $F \eta=10$ and 
$E_0/\Delta m^2 =1.73\cdot 10^{29}~{\rm eV^{-1}}$.
} 
\label{fig:fig2b} 
\end{figure}

\begin{figure}[hbt]
\begin{center}
\epsfig{file=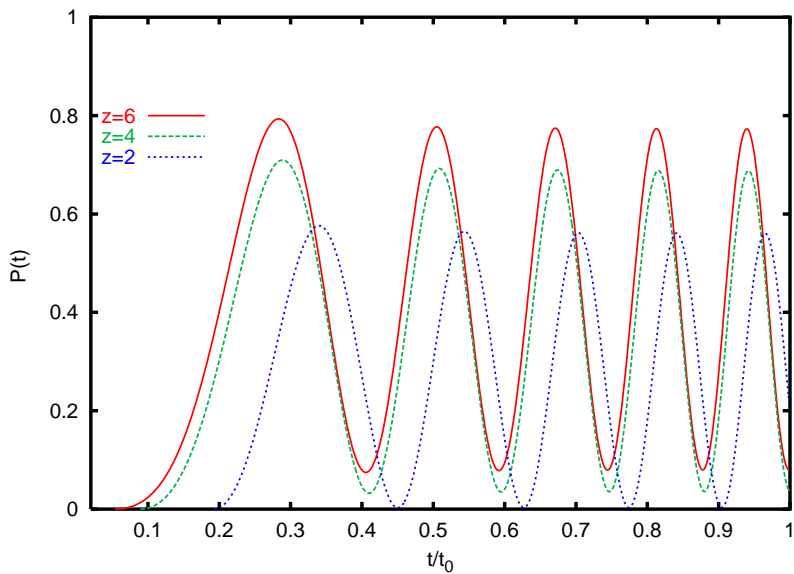, width=11truecm}
\end{center}
\caption{The same as fig. \ref{fig:fig2a} for the regime of strong breaking of adiabaticity. Here
$\sin^2 2\theta=0.5$, $F \eta=6$ and 
$E_0/\Delta m^2 =4.6\cdot 10^{30}~{\rm eV^{-1}}$. 
} 
\label{fig:fig2c} 
\end{figure}

The figures \ref{fig:fig2b}-\ref{fig:fig2c} show similar sets of curves in the cases of moderate and strong violation of adiabaticity. 

Let us consider the properties of the conversion probability $P(\nu_\alpha \rightarrow \nu_s)$ for neutrinos 
 produced at  epoch $z$ and arriving at Earth at the present epoch, $z=0$. The probability $P$ depends on $z$, on the product $F \eta$, on the  energy and mass squared difference in the ratio $E_0/\Delta m^2$, and on the mixing angle $\theta$: $P=P(z,F \eta,E_0/\Delta m^2,\theta)$.  As follows from figs. \ref{fig:fig2}-\ref{fig:fig2c}
the probability is a rapidly oscillating function of  $z$, and also of $E_0/\Delta m^2$.  
We averaged $P$ over the energy resolution interval $\Delta E_0\simeq E_0$ according to eq. (\ref{eq:aver}).
The interpretation of the numerical results can be easily given using the $z$ - $F \eta$ diagram of fig.  \ref{fig:fig1}.

\begin{figure}[hbt]
\begin{center}
\epsfig{file=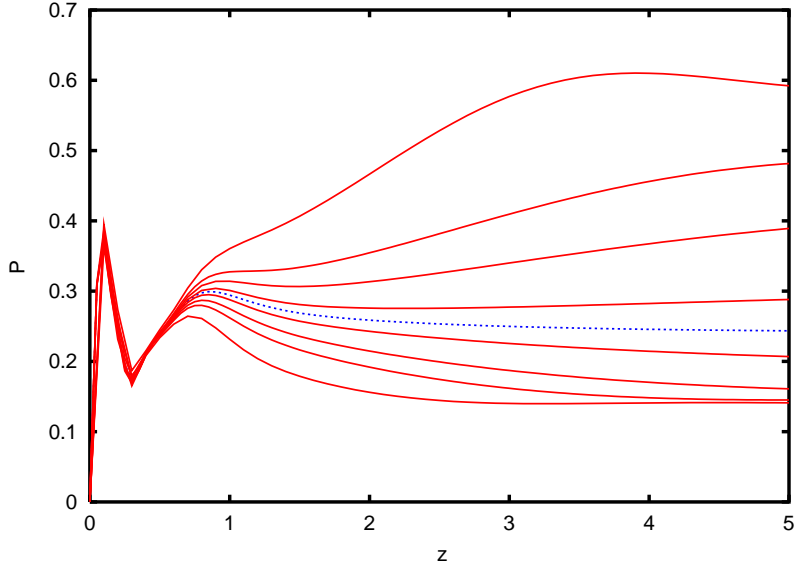, width=11truecm}
\end{center}
\caption{The $\nu_\alpha - \nu_s$  conversion probability $P$ as a function of the production epoch $z$ for various values of  $F \eta$. From the upper to the lower curve:
$F \eta=20,10,6,2,0,-2,-6,-10,-20$; the dotted line represents the vacuum oscillations probability ($F \eta=0$).
We have taken $\sin^2 2\theta=0.5$ and $E_0/\Delta m^2 =10^{31}~{\rm eV^{-1}}$. 
} 
\label{fig:fig4} 
\end{figure}

In fig.  \ref{fig:fig4} we show the dependence of the conversion probability on the production epoch $z$ for different values of $F \eta$ and fixed $E_0/\Delta m^2$ and $\sin^2 2\theta$.  The curves with  $F \eta>0$ represent the resonance channel.
For  $z\lta 1$ both vacuum oscillations and matter conversion probabilities have  oscillating behaviour. For $z\gta 2$ oscillations are averaged out, so that  the vacuum oscillation probability converges to $\sin^2 2\theta/2$~\footnote{
Notice that partial averaging exists already at small $z$ due to our integration over $\Delta E_0$. For this reason $P$ does not reach its maximal possible value $P_{max}= \sin^2 2\theta$.}.
A substantial ($\sim 10\%$) deviation from the vacuum oscillation probability due to matter effect starts at $z\simeq 1$ for  $F \eta \simeq 10$ and at $z\simeq 3$ for  $F \eta \simeq 2$. 

For $F \eta \simeq 6-10$ and $z\simeq 4-5$ neutrinos are produced at densities much higher than the resonance density and they cross the resonance at $z= 2-2.5$. The adiabaticity is broken in the resonance, however above the resonance the propagation can be adiabatic. For higher asymmetry,  $F \eta \gta 10$, the adiabaticity starts to be broken near the resonance, so that the original flavour state $\nu_\alpha\simeq \nu_{2 m}$ will evolve to $ \nu^R_{2 m}\simeq (\nu_\alpha+\nu_s)/\sqrt{2}$.  Thus, we have 
$P\simeq 1/2$. 
With the decrease of $z$ the initial state will deviate from  $\nu_{2 m}$ and the conversion probability becomes smaller.  
With the decrease of $F \eta$ the adiabaticity starts to be violated earlier (before resonance), so that the transition probability decreases.

For negative values of $F \eta$ (or for antineutrinos) the matter effect suppresses the mixing and, in consequence, the conversion effect.  However, the suppression effect is weaker than the enhancement in the resonant channel. 

Notice that for $F \eta\simeq 10$ and $z\simeq 5$ the matter effect can change the vacuum oscillation probability $P_v$  by a factor of $2$:
\begin{equation}
(P-P_v)/P_v\simeq 1~.
\label{eq:doubl}
\end{equation} 
For $z\simeq 2$ and $F \eta\simeq 10$ the deviation can reach $\sim  40\%$ and it equals $\sim 20\%$ for $F \eta\simeq 2$.
\\

\begin{figure}[hbt]
\begin{center}
\epsfig{file=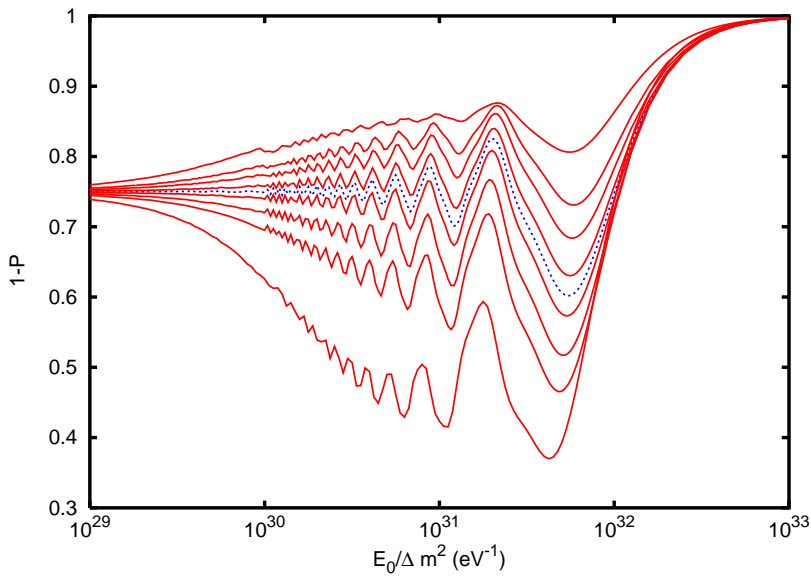, width=11truecm}
\end{center}
\caption{The survival probability $1-P(\nu_\alpha-\nu_s)$ as a function of the ratio $E_0/\Delta m^2 $ for various values of  $F \eta$. 
From the upper to the lower curve: 
$F \eta=-20,-10,-6,-2,0,2,6,10,20$;  the dotted line represents the effect of vacuum oscillations ($F \eta=0$).
We have taken $\sin^2 2\theta=0.5$ and production epoch $z=3$. 
} 
\label{fig:fig5} 
\end{figure}

In fig. \ref{fig:fig5} we show the dependence of the survival probability, $1-P$, on 
$E_0/\Delta m^2$ for production epoch $z=3$, $\sin^2 2\theta=0.5$ and various values of  $F \eta$.   
Oscillations are averaged for $E_0/\Delta m^2\lta 3\cdot 10^{30}~{\rm eV^{-1}}$; the averaging disappears at  $E_0/\Delta m^2\sim 10^{32}~{\rm eV^{-1}}$, when the oscillation length approaches the size of the horizon (see also sect. 
\ref{subsec:new.2} and fig. \ref{fig:fig8new}). 

\begin{figure}[hbt]
\begin{center}
\epsfig{file=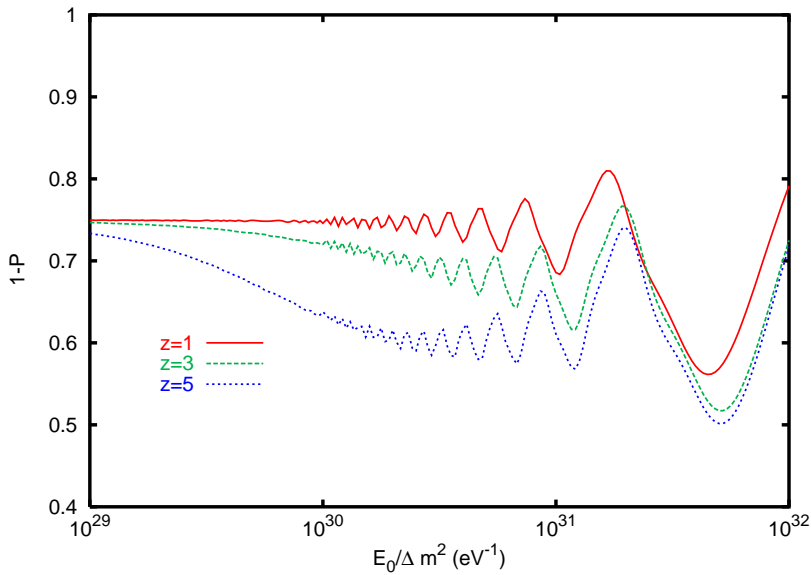, width=11truecm}
\end{center}
\caption{The survival probability $1-P(\nu_\alpha-\nu_s)$ as a function of the ratio $E_0/\Delta m^2 $ for various values of  the production epoch $z$.  We have taken $\sin^2 2\theta=0.5$ and  $F \eta=6$.
} 
\label{fig:fig6} 
\end{figure} 

The matter effect increases with $E_0/\Delta m^2$. For  $E_0/\Delta m^2\lta 5\cdot 10^{30}~{\rm eV^{-1}}$ the resonance epoch $z_R$ (see eq. (\ref{eq:zr})) is earlier than the production epoch of the neutrinos. Thus the neutrinos do not cross the resonance and the matter effects is realized mainly in the epoch of neutrino production, when the potential (\ref{eq:potefuniv}) was larger (see fig.  \ref{fig:fig2}). The value of the effect is determined by the mixing in matter at the production time.  With the increase of  $E_0/\Delta m^2$ the resonance epoch $z_R$ approaches the production epoch  (see fig.  \ref{fig:fig1}). As a consequence the mixing at production, and therefore the matter effect, increase.   
The maximal matter effect is achieved at energies for which the resonance condition is fulfilled at production epoch or slightly later (notice that the adiabaticity is strongly broken at resonance).  For $z\simeq 3$ this occurs in the interval  $E_0/\Delta m^2\simeq  10^{31}-10^{32}~{\rm eV^{-1}}$.  For $z\simeq 5$ maximal matter effect is realized at $E_0/\Delta m^2\simeq  (5-7)\cdot 10^{30}~{\rm eV^{-1}}$ (fig. \ref{fig:fig6}). 
\\

\begin{figure}[hbt]
\begin{center}
\epsfig{file=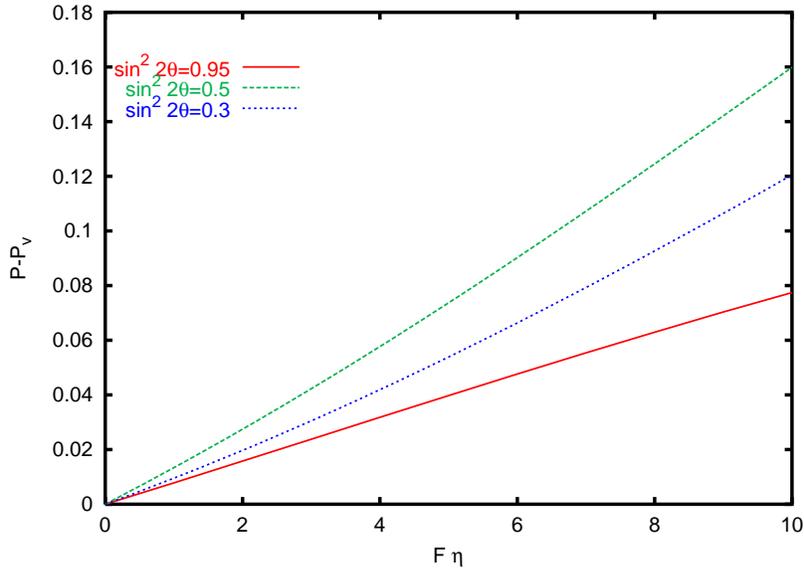, width=11truecm}
\end{center}
\caption{The deviation with respect to the vacuum oscillation probability, $P(\nu_\alpha-\nu_s)-P_v$ as a function of the product $F \eta$ for various values of $\sin^2 2\theta$. We have taken production epoch $z=3$ and $E_0/\Delta m^2 =10^{31}~{\rm eV^{-1}}$. 
} 
\label{fig:fig7} 
\end{figure}

In fig. \ref{fig:fig7} we show the dependence of the matter effect, i.e. the difference $P-P_v$, on the quantity $F \eta$ for various values of the mixing angle.  For the parameters used in the plot the neutrinos are produced close to the resonance and the adiabaticity is strongly violated in the resonance.  The matter effect can be estimated as the deviation of the jump probability from 1: 
\begin{equation}
1-P_{LZ}\simeq 1-\exp({-\pi \chi_R/2})~.
\label{eq:jump}
\end{equation}
  In our case $\chi_R\ll 1$, so that the matter effect is proportional to  $F \eta$:
\begin{equation}
P-P_v\simeq {\pi \over 2} \chi_R \propto F \eta \tan^2 2\theta~,
\label{eq:linbeh}
\end{equation}
according to eq. (\ref{eq:adiabresnum}). This explains the linear increase of the matter effect with $\eta$ and $F$.
\\

\begin{figure}[hbt]
\begin{center}
\epsfig{file=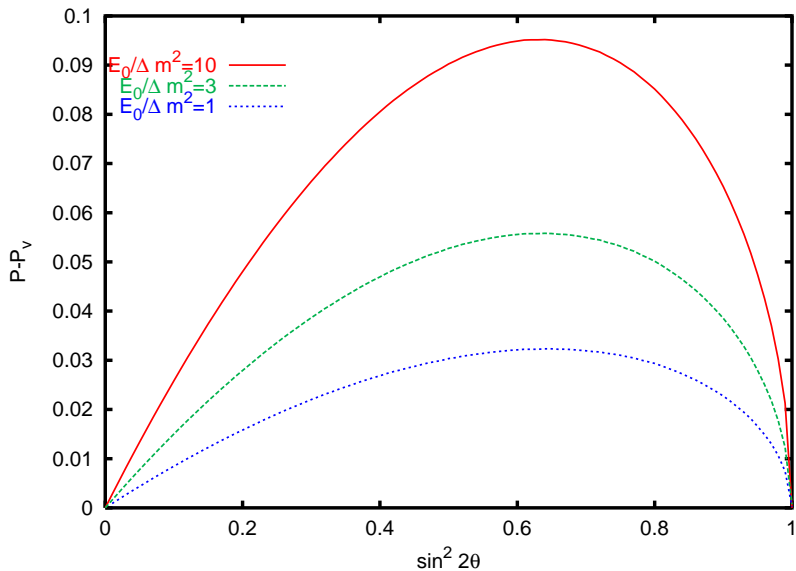, width=11truecm}
\end{center}
\caption{The deviation with respect to the vacuum oscillation probability, $P(\nu_\alpha-\nu_s)-P_v$ as a function of  $\sin^2 2\theta$, for various values of $E_0/\Delta m^2 $ (in units of $10^{30}~{\rm eV^{-1}}$). We have taken production epoch $z=3$ and $F \eta=6$.
} 
\label{fig:fig8} 
\end{figure}

In  fig. \ref{fig:fig8} we show the dependence of $P-P_v$ on the mixing parameter $\sin^2 2 \theta$ for different values of the ratio $E_0/\Delta m^2$ and fixed production epoch $z=3$ and $F \eta=6$. The neutrinos are produced in the resonance epoch or after it depending on their energy. For small mixing the matter effect is proportional to the mixing parameter $\sin^2 2 \theta_m$ at the production time. This explains the linear increase of the effect with $\sin^2 2 \theta$ ($\sin^2 2 \theta_m\propto \sin^2 2 \theta$) and with $E_0/\Delta m^2$ (for $E_0/\Delta m^2\sim 10^{31}~{\rm eV^{-1}}$ the production epoch coincides with the resonance one).
For maximal mixing, $\sin^2 2 \theta=1$, the average probability takes the value $P=1/2$ independently on adiabaticity violation \cite{Gonzalez-Garcia:2000ve}. Therefore in this case $P-P_v=0$. The maximum deviation from vacuum  oscillation effect is realized at $\sin^2 2 \theta\simeq 0.65$.
\\


\section{Conversion effects on diffuse neutrino fluxes}
\label{sec:4}

The results we have discussed in the sections \ref{subsec:new.2} and \ref{subsec:3.3} 
describe the conversion effect for a beam of neutrinos produced by a single 
source at a certain epoch
$z$.   Presently, the possibilities of detection of neutrinos from single
sources are limited to objects with redshift $z\ll 1$. For these neutrinos no substantial effect is expected\footnote{Some effect can appear due to conversion in halos  of galaxies and of clusters of galaxies \cite{next}.}. There is a hope, however, to detect the diffuse (integrated) neutrino flux which is produced by all the cosmological sources.  For this flux matter effects can be observable.

In what follows we will calculate the ratio $F^\alpha(E_0)/F^\alpha_0(E_0)$, where  $F^\alpha(E_0)$ and $F^\alpha_0(E_0)$ are the present diffuse fluxes 
of neutrinos of  given flavour, $\nu_\alpha$, and a given energy, $E_0$, with and without conversion.
The ratio can be written as:
\begin{eqnarray} 
{F^\alpha(E_0)\over F^\alpha_0(E_0)}=1-\bar{P}_\alpha(E_0)~,
\label{eq:depl}
\end{eqnarray} 
where $\bar{P}_\alpha$ is the averaged transition probability:
\begin{eqnarray}
\bar{P}_\alpha(E_0)\equiv {1\over F^\alpha_0(E_0)}
\int^{z_{max}}_{0} {dF^\alpha_0(E_0,z)\over dz} P_\alpha(E_0,z) dz ~. 
\label{eq:convolution}
\end{eqnarray} 
Here $P_\alpha(E_0,z)$ is the transition probability for neutrinos produced in the epoch $z$, which has been discussed in sections \ref{subsec:new.2} and \ref{subsec:3.3}. The quantity ${dF^\alpha_0(E_0,z)}$ is the contribution of the neutrinos  $\nu_\alpha$ produced in the interval $[z,z+dz]$ to the present flux in absence of oscillations. 

We first derive the general expression for the differential flux $dF^\alpha_0(E_0,z)$.  Let $f(E)$ be the flux of neutrinos generated by a single source. Then the total number of neutrinos produced in the unit volume in the time interval $[t, t+dt]$ with energy in the interval $[E,E+dE]$ can be written as:
\begin{equation}
f(E) n(t) dE dt~,
\label{eq:nuuniv}
\end{equation}
where $n(t)$ is the concentration of sources  in the epoch $t$. The contribution of these neutrinos to the present flux equals: 
\begin{equation}
dF^\alpha_0(E_0,z)={c \over 4\pi}f(E) n(t) (1+z)^{-3} {dE \over dE_0} dt~,
\label{eq:dfgen}
\end{equation}
where $c$ is the speed of light and  the factor $(1+z)^{-3}$ accounts for the expanding volume of the universe.
Transferring from $t$ to $z$-variable we get: 
\begin{eqnarray} 
dF^\alpha_0(E_0,z)={3 c  t_0\over 8\pi} f(E) n(z) (1+z)^{-11/2} {dE \over dE_0} dz~.
\label{eq:fluxxdec}
\end{eqnarray}
The relation between the energy $E$ and the present neutrino energy $E_0$ includes, in general, effects of energy losses and of redshift. Neglecting absorption we have $dE/dE_0=(1+z)$. 
The density of sources, $n(z)$, can be expressed in terms of the comoving density $n_c$ as $n(z)=(1+z)^3 n_c(z)$.  Notice that $n_c=const$, if the number of sources in the universe is constant in time. Thus, the evolution of sources is described by the dependence of $n_c$ on the redshift $z$.

In terms of $n_c$ and $E_0$ we get finally:
\begin{eqnarray} 
dF^\alpha_0(E_0,z)={3 c  t_0\over 8\pi} f(E_0(1+z)) n_c(z) (1+z)^{-3/2} dz~.
\label{eq:fluxfinal}
\end{eqnarray}

Inserting $dF^\alpha_0(E_0,z)$ in eq. (\ref{eq:convolution}) we find:
\begin{eqnarray}
\bar{P}_\alpha(E_0)= {1\over F^\alpha_0(E_0)}{3 c  t_0\over 8\pi}\int  f(E_0(1+z)) n_c(z) (1+z)^{-3/2} P_\alpha(E_0,z) dz~,
\label{eq:convgrb}
\end{eqnarray} 
and $F^\alpha_0(E_0)$ is given by the same expression with $P_\alpha=1$.

In what follows we will calculate the survival probability $1-\bar{P}_\alpha$ for various possible sources of high-energy neutrinos, assuming certain forms for the produced flux $f(E)$ and  the concentration of sources $n_c$.


\subsection{Conversion of neutrinos from AGN and GRBs}
\label{subsec:4.1}

There is an evidence that cosmological sources like GRBs and AGN were more numerous in the past. In particular, the density of 
GRBs evolved as \cite{Hogg:1998wi}:
\begin{eqnarray}
n_c(z)\propto \cases {(1+z)^{3}  & $z\leq z_p$  \cr
                      (1+z_p)^3  &  $z_p<z\leq z_{max}$ \cr 
                       \sim 0    & $z>z_{max}$ \cr }~.
\label{eq:grbdistr}  
\end{eqnarray}
where $z_p$ is estimated to be $z_p\simeq 1-2$ \cite{Hogg:1998wi}.  
The energy spectrum of neutrinos from GRBs scales as a power law \cite{Waxman:1997hj} : 
\begin{eqnarray} 
f(E)\propto {1\over E^2}={1\over {E^2_0 (1+z)^2}}~.
\label{eq:pwl}
\end{eqnarray}
Combining eqs. (\ref{eq:grbdistr}) and (\ref{eq:pwl}) with (\ref{eq:convgrb})  we find the averaged probability:
\begin{eqnarray}
\bar{P}_\alpha(E_0)={1\over N_p}\left[ \int_{0}^{z_p} (1+z)^{-1/2} P_\alpha(E_0,z) dz + \int_{z_p}^{z_{max}} (1+z)^{-7/2} P_\alpha(E_0,z) dz \right]~,
\label{eq:grb}  
\end{eqnarray} 
where  the normalization factor $N_p$ is given by the expression in square brackets with $P_\alpha=1$.  According to eq. (\ref{eq:grb}) the contribution of the recent epochs to the present flux  is enhanced in spite of the the larger number of sources in the past. This leads to suppression of the matter effects, which are more important at large $z$.  

\begin{figure}[hbt]
\begin{center}
\epsfig{file=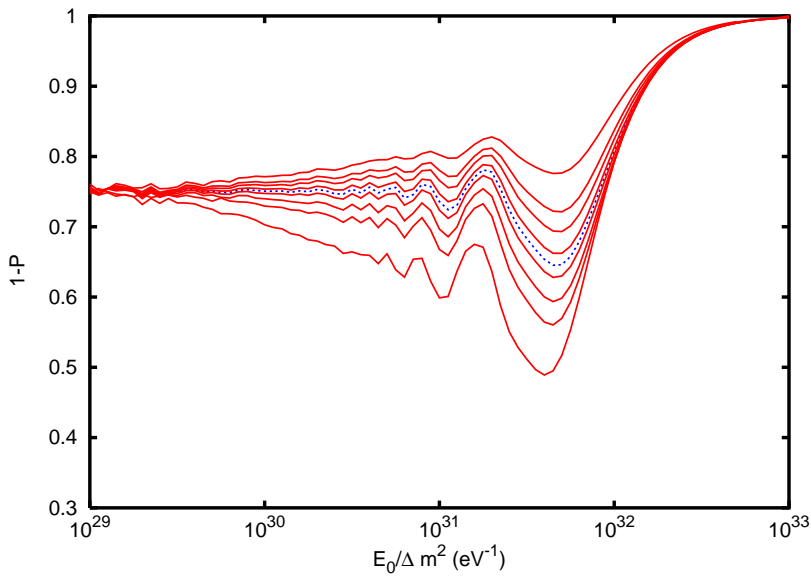, width=11truecm}
\end{center}
\caption{The averaged survival probability for $\nu_\alpha-\nu_s$  channel, $1-\bar{P}_\alpha$, as a function of the ratio $E_0/\Delta m^2 $ for the diffuse flux of neutrinos from GRBs. The curves correspond to various values of  $F \eta$.
From the upper to the lower curve: 
$F \eta=-20,-10,-6,-2,0,2,6,10,20$; the dotted line represents the effect of vacuum oscillations ($F \eta=0$).
We have taken $\sin^2 2\theta=0.5$.
} 
\label{fig:fig9} 
\end{figure}

The figure \ref{fig:fig9} shows the averaged survival probability , 
$1-\bar{P}_\alpha$, for $\nu_\alpha-\nu_s$ conversion channel,  as a function of $E_0/\Delta m^2$ for different values 
of $F \eta$. We have taken $z_p=2$ and $z_{max}=5$. 
The averaged probability is rather close to the non-averaged one (see fig. \ref{fig:fig4}) for neutrinos produced at $z\simeq z_p=2$. Indeed, the contribution to the flux from the earlier epochs,   $z\gta z_p$ is strongly suppressed, according to eq. (\ref{eq:grb}). The integration over $z$ leads to some smoothing of the oscillatory behaviour of the probability.  The deviation of the ratio $F^\alpha(E_0)/F^\alpha_0(E_0)$ from its vacuum oscillation value can reach $\sim 25\%$. Maximal effect is realized for $F \eta\simeq 20$ in the resonance interval  $E_0/\Delta m^2\sim (1-5)\cdot 10^{31}~{\rm eV^{-1}}$.  For  $F \eta\simeq 2$ the effect is about $ (3-4)\%$.

\begin{figure}[hbt]\begin{center}
\epsfig{file=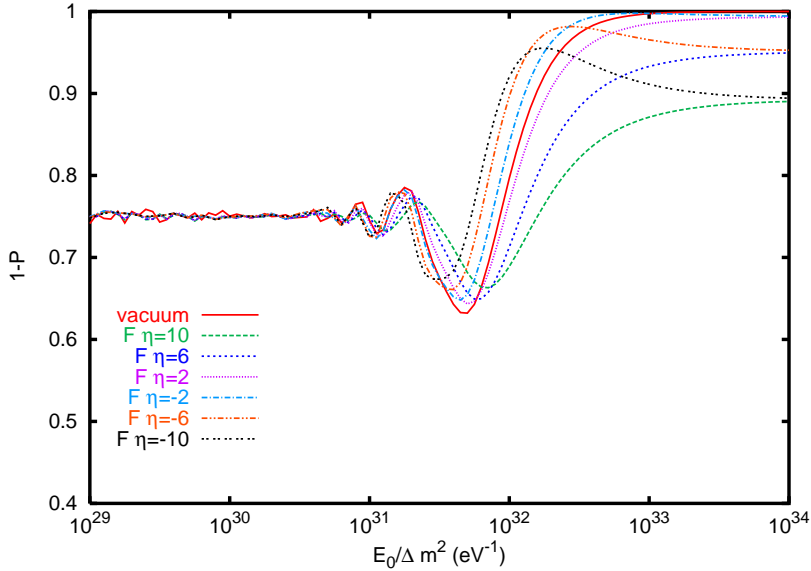, width=11truecm}
\end{center}
\caption{The averaged survival probability for  $\nu_\alpha-\nu_\beta$  oscillations, $1-\bar{P}_\alpha$, as a function of the ratio $E_0/\Delta m^2 $ for the diffuse flux of neutrinos from GRBs. The curves correspond to various values of  $F \eta$.
We have taken $\sin^2 2\theta=0.5$.
} 
\label{fig:fig9new} 
\end{figure}

For conversion between active flavours the results are  shown in fig. \ref{fig:fig9new}: the deviation of the survival probability  from the value given by vacuum oscillations can be as large as $\sim 10\%$  for large asymmetry, $F\eta\gta 10$, and high energies, $E_0/\Delta m^2\gta  10^{32}~{\rm eV^{-1}}$, for which the matter-induced oscillation phase $\Phi_{matt}$ dominates over the vacuum oscillation phase $\Phi_{vac}$.

The astrophysical data about AGN indicate that the distribution of these objects has maximum at $z\sim 2$ \cite{Miyaji:1999be}, with a rapid decrease of the concentration with $z$. The power law $f(E)\propto E^{-2}$ is a good approximation for the most energetic part of the spectrum \cite{Stecker:1996th}.
For these reasons, in the case of AGN the results are similar to those discussed here for neutrinos from GRBs.


\subsection{Conversion of neutrinos from heavy particle decay} 
\label{subsec:4.2} 
Very heavy particles, with mass of the order of the
grand unification scale, are supposed to be produced in the
universe by topological defects, e.g. in monopole-antimonopole annihilation, 
cosmic strings evaporation, etc. \cite{Bhattacharjee:1997ps}.  These particles would then decay very quickly, 
with lifetime
$\tau\ll t_0$, into leptons and hadrons.  Neutrinos may be produced either
directly, as primary decay products, and/or as secondary products from decays of hadrons.

Let us calculate the contribution of the neutrinos produced in the epoch $z$ to the present flux: $dF^\alpha_0(E_0,z)/dz$. In assumption of very fast decay of the heavy particle, $X$ (so that the production epochs of $X$ and of the neutrinos coincide), we can write the total number of neutrinos produced in the unit volume in the time interval $[t, t+dt]$ with energy in the interval $[E, E+dE]$ as:
\begin{eqnarray} 
{dn_X(t)\over dt}{dN_\nu\over dE}dt dE~,
\label{eq:phi}
\end{eqnarray}
where $dn_X(t)$ is the number of $X$ particles produced in the interval $[t, t+dt]$  in the unit volume, and $dN_\nu$ is the number of neutrinos in the energy interval  $[E, E+dE]$ produced  by a single particle $X$.
The contribution of the neutrinos produced in the epoch $t$, eq. (\ref{eq:phi}), to the present neutrino flux is:
\begin{equation}
dF^\alpha_0(E_0,z)={c \over 4\pi}{dn_X(t)\over dt}{dN_\nu\over dE} 
(1+z)^{-3} {dE \over dE_0} dt~,
\label{eq:dfx}
\end{equation}
where we have taken into account the expansion of the universe. In terms of the redshift $z$ we get:
\begin{eqnarray} 
dF^\alpha_0(E_0,z)={3 c  t_0\over 8\pi} {dn_X\over dt}(z){dN_\nu\over dE}(E_0(1+z)) (1+z)^{-9/2} dz~,
\label{eq:xdec}
\end{eqnarray}
where we used also the relation $E=E_0(1+z)$. 

The production rate of the $X$ particles can be written as:
\begin{eqnarray} 
{dn_X(t)\over dt}\propto t^{-4+p}\propto (1+z)^{6-{3\over 2}p}~,
\label{eq:prodrate}
\end{eqnarray}
where  $p=1$
for monopole-antimonopole annihilation and cosmic strings \cite{Bhattacharjee:1990js} and  $p=2$
for constant comoving production rate.

For the fragmentation function of neutrinos we take a power law:
\begin{eqnarray} 
{dN_\nu\over dE}\propto E^{\alpha}=E^{\alpha}_0 (1+z)^{\alpha}~.
\label{eq:fragf}
\end{eqnarray}
If the neutrinos are produced mainly by hadronic decays
 the fragmentation function has a  polynomial form \cite{Wichoski:1998kh}. 
The leading term of the polynome gives  the expression (\ref{eq:fragf}) with $\alpha=-3/2$. 

Inserting the  expressions from (\ref{eq:xdec}), (\ref{eq:prodrate}) and (\ref{eq:fragf}) in eq. (\ref{eq:convolution})  we get:
\begin{eqnarray} 
\bar{P}_\alpha(E_0)={1 \over N_p} \int(1+z)^{6-{3\over 2}p+\alpha}  P_\alpha(E_0,z)dz~,
\label{eq:phixdec}
\end{eqnarray} 
and,  
for $p=1$ and $\alpha=-3/2$:
\begin{eqnarray} 
\bar{P}_\alpha(E_0)={1 \over N_p} \int (1+z)^{-{3\over 2}} P_\alpha(E_0,z)dz ~.
\label{eq:fxd}
\end{eqnarray}
Here $N_p$ is a normalization factor.

We perform
the integration (\ref{eq:fxd}) starting from the absorption epoch $z_{abs}$.  The contribution  of the neutrino flux produced at $z\gta z_{abs}$ is very small due to absorption\footnote{ Clearly, the energy of the neutrinos at production can not exceed the mass of the parent particle, $X$. This gives 
a further constraint on the upper
integration limit: $1+z_{max}\lta {m_X/E_0}$. Stronger bounds can be found in some specific production mechanisms: taking, for instance, $X\rightarrow \pi^{+} \pi^{-}$ and subsequent production of neutrinos by pion decay, one gets
$1+z_{max}\lta 0.2134 {m_X/E_0}$ \cite{Bhattacharjee:1992zm}.  For $E_0\lta 10^{22}~{\rm eV}$ and 
$m_X\sim 10^{16}~{\rm GeV}$ this gives the constraint $z_{max}\lta 200$, which is weaker
than the one given by absorption.}.
The dominant absorption processes are $\nu-\nu$ and $\nu-\bar{\nu}$ interaction
with the neutrino background. The absorption epoch is given by:
\begin{eqnarray}
1+z_{abs}=\left[ {d_{abs}\over d_U}+1\right]^{2\over 3}~,
\label{eq:zabs}
\end{eqnarray}
where $d_U$ is given in eq. (\ref{eq:du}) and $d_{abs}$ is the 
the absorption width, which depends on the $\nu-\nu$ energy squared in the center of
mass, $s_Z$, (see eq. (\ref{eq:param})).
Taking, for instance, $E\lta 10^{22}~{\rm eV}$ and $m_\nu\lta 0.05~{\rm eV}$, 
we have $s_Z\lta 0.1$, and the corresponding absorption width
is $d_{abs}\gta 1.5\cdot 10^{34}~{\rm cm^{-2}}$ \cite{Roulet:1993pz}. With this value and
$\eta\simeq 10$ eq. (\ref{eq:zabs}) gives $z_{abs}\simeq 50$. 

\begin{figure}[hbt]
\begin{center}
\epsfig{file=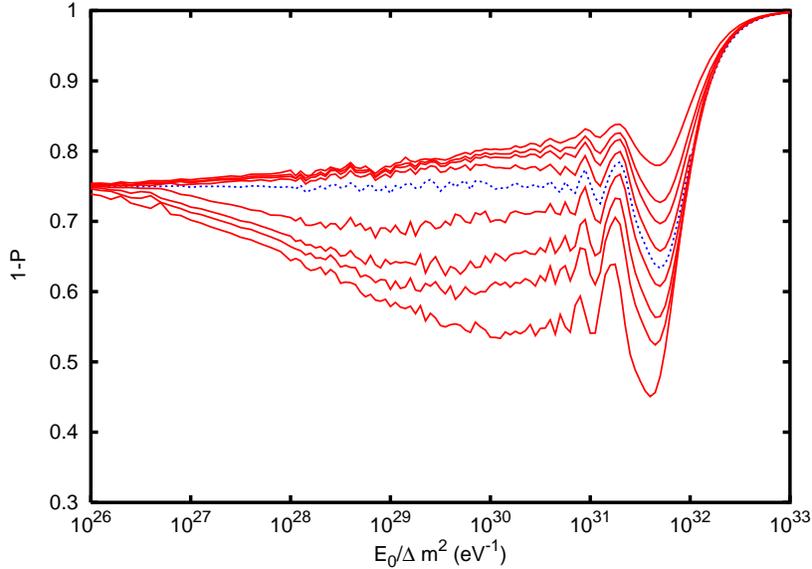, width=11truecm}
\end{center}
\caption{The averaged survival probability for $\nu_\alpha-\nu_s$  channel, $1-\bar{P}_\alpha$, as a function of the ratio $E_0/\Delta m^2 $ for  neutrinos from the decay of heavy relics. The curves correspond to various values of  $F \eta$. 
From the upper to the lower curve: 
$F \eta=-20,-10,-6,-2,0,2,6,10,20$; the dotted line represents the effect of vacuum oscillations  ($F \eta=0$).
We have taken $\sin^2 2\theta=0.5$.
} 
\label{fig:fig10} 
\end{figure}

In the figure \ref{fig:fig10} we show the averaged survival probability $1-\bar{P}_\alpha$ for $\nu_\alpha-\nu_s$ conversion channel, as a function of $E_0/\Delta m^2$ for different values of $F \eta$. 
One can see that, in contrast with the case of neutrinos from GRBs, the deviation of the ratio $F^\alpha(E_0)/F^\alpha_0(E_0)$ from the value given by vacuum oscillation is significant (larger than $\sim 10\%$) in a wide range of energies: $E_0/\Delta m^2\simeq 10^{26}-10^{32}~{\rm eV^{-1}}$.  

For a given value of $E_0/\Delta m^2$ the matter effects are determined by the corresponding resonance epoch, $z_R$, and adiabaticity in resonance.
For $E_0/\Delta m^2\lta 10^{29}~{\rm eV^{-1}}$ the resonance was realized at $z_R\gta 10$, when the adiabaticity condition was fulfilled  (see fig. \ref{fig:fig1}). Therefore, the matter effects are dominated by resonant adiabatic conversion which occurs for neutrinos produced at $z>z_R\sim 10$.  As discussed in sect. \ref{subsec:3.3}, these neutrinos undergo almost total conversion (see fig. \ref{fig:fig2}), however, their contribution to $\bar{P}_\alpha$ is suppressed according to eq. (\ref{eq:fxd}).

For $E_0/\Delta m^2\gta 10^{29}~{\rm eV^{-1}}$ the resonance is realized at $z_R\lta 10$, when the adiabaticity is broken  (fig. \ref{fig:fig1}), so that the matter effect is mostly due to non-adiabatic conversion and oscillations in the production epoch.

The maximal effect is realized in the interval $E_0/\Delta m^2\simeq 10^{29}-5\cdot  10^{31}~{\rm eV^{-1}}$; the relative deviation of $F^\alpha(E_0)/F^\alpha_0(E_0)$ with respect to the vacuum oscillations value equals $\sim 10\%$ for $F \eta=2$ and can be as large as $50\%$ for $F \eta=20$.
 
\begin{figure}[hbt]
\begin{center}
\epsfig{file=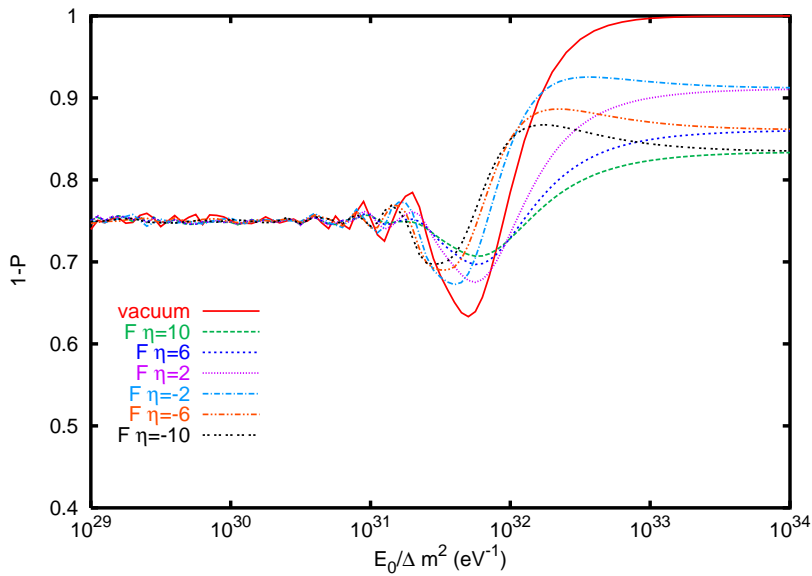, width=11truecm}
\end{center}
\caption{The averaged survival probability for  $\nu_\alpha-\nu_\beta$  oscillations, $1-\bar{P}_\alpha$, as a function of the ratio $E_0/\Delta m^2 $ for the diffuse flux of neutrinos from the decay of heavy relics. The curves correspond to various values of  $F \eta$.
We have taken $\sin^2 2\theta=0.5$.
} 
\label{fig:fig9bis} 
\end{figure}

The figure \ref{fig:fig9bis} shows the average survival probability, $1-\bar{P}_\alpha$, for active-active conversion.  
We see that, similarly to what discussed for neutrino from GRBs, a substantial ($\sim 15\%$) matter effect requires large asymmetry, $F\eta\gta 10$, and very high energies, $E_0/\Delta m^2\gta 10^{32}~{\rm eV^{-1}}$, for which the matter contribution to the oscillation phase is dominant.


\section{Observable effects}
\label{subsec:4.3}
Let us consider the experimental signatures of matter effects on neutrino 
propagation. The observable effects depend on the specific scheme of neutrino masses and mixings and on the initial flavour composition of the neutrino flux.


\subsection{Conversion of cosmic neutrinos and neutrino mass schemes}
\label{subsec:5.1}

As follows from the analysis of sections \ref{subsec:4.1}-\ref{subsec:4.2},
a significant matter effect on active-active oscillations of high-energy neutrinos requires:
\begin{eqnarray}
{E_0 \over \Delta m^2}\gta 10^{32}~{\rm eV^{-1}}~. 
\label{eq:requirenaa}
\end{eqnarray}
This is the condition for  which the matter-induced oscillation phase, $\Phi_{matt}$, dominates over the vacuum one, $\Phi_{vac}$ (see section \ref{subsec:new.2}). 
For conversion into a sterile neutrino the matter effect is substantial in the ranges:
\begin{eqnarray}  
{E_0 \over \Delta m^2}\gta \cases {10^{30}~{\rm eV^{-1}}  & for AGN, GRBs  \cr
                      10^{28}~{\rm eV^{-1}}  &  for heavy relics decay~. \cr }
\label{eq:requiren}
\end{eqnarray}
For $E_0\lta 10^{21}~{\rm eV}$, the conditions (\ref{eq:requirenaa})-(\ref{eq:requiren}) imply: 
\begin{eqnarray}
\Delta m^2\lta  \cases {
10^{-11}~{\rm eV^{2}}  & for $\nu_\alpha-\nu_\beta$  \cr
10^{-7}~{\rm eV^{2}} &  for $\nu_\alpha-\nu_s$~. \cr }
\label{eq:requirm}
\end{eqnarray}
For both the active-active and active-sterile channels the mixing angle should be large enough and, for $\nu_\alpha-\nu_s$, not too close to maximal (see fig. \ref{fig:fig8}):
\begin{eqnarray} 
0.1\lta \sin^2 2\theta\lta 0.95~.
\label{eq:requirmix}
\end{eqnarray}
In the three neutrino schemes which explain the solar and atmospheric neutrino data the effect can be realized for $\nu_e$-$ \nu_\tau/\nu_\mu$ mixing and the vacuum oscillation (VO) solution of the solar neutrino problem. 
If the LMA, the SMA  or the LOW  solution are confirmed, the effect of medium on vacuum oscillations of cosmic neutrinos can be neglected.

In presence of a sterile neutrino the conditions  (\ref{eq:requiren}) - (\ref{eq:requirmix}) can be realized in a number of situations.  
Oscillations of electron neutrino into a sterile state, $\nu_e-\nu_s$, with $\Delta m^2\lta 10^{-11}~{\rm eV^{2}}$ and mixing close to maximal represent a possible solution of the solar neutrino problem \cite{Bahcall:2001hv}.
Another possibility is to consider, e.g., the hierarchical mass spectrum with $m_3\sim \sqrt{\Delta m_{atm}^2}$, $m_2\sim \sqrt{\Delta m_{\odot}^2}$,  $m_1\lta  10^{-4}~{\rm eV}$ and $m_0< m_1$
so that  $\Delta m^2_{1 0}\simeq m^2_1\lta  10^{-7}~{\rm eV^{2}}$.
In the simplest case the sterile state is mixed only in the lightest mass eigenstates $\nu_1$ and $\nu_0$.  
The mixing angle is only weakly restricted by the solar neutrino data\footnote{
No restriction exists for $\Delta m_{1 0}^2\ll 10^{-11}~{\rm eV^{2}}$; bounds follow from the solar neutrino data for $\Delta m_{1 0}^2\gta 10^{-11}~{\rm eV^{2}}$ (in this case the solar neutrino data should be treated in three neutrino context).}.

\subsection{Flavour composition of detected fluxes}
\label{subsec:5.2}

Let us consider the numbers of events $N_\alpha$ and $N^0_\alpha$ induced in a detector by neutrinos of different flavours $\alpha$ with and without conversion respectively. These quantities are determined by the present fluxes 
$F^\alpha$ and $F^\alpha_0$ (see section \ref{sec:4}); 
if the detector provides total energy reconstruction and optimal event selection, the flavour composition of the numbers of  detected events coincides with that of the fluxes.  

In what follows we consider two possible types of flavour composition for the numbers of events in absence of conversion:
\\
\noindent
1). CP-symmetric: 
$N^0_{\alpha}=N^0_{\bar{\alpha}}$ ($\alpha=e,\mu,\tau$).  
As far as flavour content is concerned we take the normalized numbers of events:
\begin{eqnarray} 
&&(N^0_{e},N^0_{\mu},N^0_{\tau})=(1,2,0)\nonumber\\
&&(N^0_{\bar{e}},N^0_{\bar{\mu}},N^0_{\bar{\tau}})=(1,2,0)~.
\label{eq:cpsflux}
\end{eqnarray}
Such a flavour composition is expected for neutrinos produced by the decays of $\pi^+$ and
$\pi^-$ mesons, which in turn appear in the process 
$X\rightarrow \pi^+ \pi^{-}$ (see section \ref{subsec:4.2}).
\\

\noindent
2). CP-asymmetric: $N^0_{\alpha}\neq N^0_{\bar{\alpha}}$.
We consider 
\begin{eqnarray} 
&&(N^0_{e},N^0_{\mu},N^0_{\tau})=(1,1,0) \nonumber \\
&&(N^0_{\bar{e}},N^0_{\bar{\mu}},N^0_{\bar{\tau}})=(0,1,0)~. 
\label{eq:cpaflux}
\end{eqnarray}
This flavour composition is realized for neutrinos produced by the scattering of highly energetic protons on a photon background, where the $\pi^+$ decay gives the dominant contribution. 
The $p$-$\gamma$ interaction is supposed to be the main mechanism of neutrino production in GRBs \cite{Waxman:1997hj}.
\\

Neutrinos of different flavours produced  in the same decay reaction ($X$ or $\pi$ decay) share the energy of the parent particle equally with good approximation.
Therefore the produced fluxes of neutrinos and antineutrinos of different flavours  have the same energy dependence, and, in absence of conversion, the ratios $N^0_{e}/N^0_{\mu},N^0_{e}/N^0_{\tau},N^0_{\mu}/N^0_{\tau}$ are expected to be energy-independent.

In presence of vacuum oscillations the ratios of numbers of events are approximately independent of energy in two intervals: \\
(i) $E_0/\Delta m^2\lta 5\cdot 10^{30}~{\rm eV^{-1}}$ (see figs. \ref{fig:fig9}-\ref{fig:fig10}), where oscillations are averaged out.\\
(ii)  $E_0/\Delta m^2\gta 5\cdot 10^{32}~{\rm eV^{-1}}$, where the vacuum oscillation phase is very small or, equivalently, the vacuum oscillation length exceeds the size of the horizon.  In this case the conversion probability is negligibly small and the ratios of numbers of events approach their values in absence of oscillations.\\

\noindent
The effects of vacuum oscillations are modified by the interaction with the neutrino background.
According to the results of sections \ref{subsec:4.1}-\ref{subsec:4.2} we find that:

\begin{enumerate}

\item 
The energy dependence of the ratios of numbers of events in the interval (i) would be a signal of active-sterile conversion with matter effects\footnote{If some difference exists in the energy dependences of the  original fluxes of neutrinos of different flavours, this would appear in the total number of events $N^0_{tot}=\sum_{\alpha} N^0_{{\alpha}}$, in contrast with the effect of neutrino conversion. Thus, the energy-dependence of ratios of numbers of events due to matter effects can be distinguished.}. 

\item 
The deviation of the ratios of numbers of events in the interval (ii) from the values expected in absence of conversion would indicate matter-affected active-active oscillations.  Two elements, however, will make the  identification of the effect difficult:
 its appearance at very high energies, close to the end of the predicted spectra of ultra-high energy neutrinos, and the uncertainties on the flavour composition of the neutrino fluxes at production.

\end{enumerate}

We notice an interesting aspect: the interaction with the neutrino background produces strongly different effects on active-active and active-sterile oscillations. Thus the observation of such effects would neatly distinguish between the two channels. In particular, the observation of the characteristics described in 1. would give indication of the existence of a sterile neutrino.

\subsection{Ratios of numbers of events: active neutrino mixing}
\label{subsec:5.3} 

For three neutrino flavours, $\nu_e$,
$\nu_\mu$, $\nu_\tau$,  
the relation between the numbers of events $N_\alpha$ and $N^0_\alpha$ can be expressed as:
\begin{eqnarray} 
&&\vec{N}_\nu={\cal P}\vec{N}^0_\nu \hskip 1cm ~, 
\label{eq:matrixform}
\end{eqnarray}
where:
\begin{eqnarray}
&&\vec{N}^0_\nu=(N^0_{e},N^0_{\mu},N^0_{\tau}) \nonumber\\
&&\vec{N}_\nu=(N_{e},N_{\mu},N_{\tau})~,     
\label{eq:matrixdef}
\end{eqnarray}
and ${\cal P}$ is the matrix of conversion probabilities:  
${\cal P}_{\alpha \beta}\equiv P(\nu_\alpha \rightarrow \nu_\beta)$, 
($\alpha,\beta=e,\mu,\tau$). 

As an example we consider the scenario introduced in section \ref{subsubsec:2.2.1}, in which the solar neutrino problem is solved by $\nu_e-\nu_\mu/\nu_\tau$ vacuum oscillations with $\Delta m^2_{\odot}=\Delta m^2_{2 1 }\simeq 10^{-11}~{\rm eV^{2}}$ and the atmospheric neutrino anomaly is explained by $\nu_\mu - \nu_\tau$ oscillations with $\Delta m^2_{atm}=\Delta m^2_{3 2}\simeq 10^{-3}~{\rm eV^{2}}$.  The mixing matrix is given by eq. (\ref{eq:10}).

Since the values of $\Delta m^2_{3 2}$ and $\Delta m^2_{31}$ ($\Delta m^2_{31}\simeq \Delta m^2_{32}$) are out of the range of  sensitivity to matter effects (see sect. \ref{subsec:5.1}) the the oscillations due to 
$\Delta m^2_{31}$ and $\Delta m^2_{32}$ are described by the average vacuum
oscillation probability. 
 The neutrino background influences the
$\nu_1-\nu_2$ system only.
In these specific circumstances matter effects show up in the conversion of 
$\nu_e=\cos 2\theta \nu_1+\sin 2\theta \nu_2$ into the 
orthogonal state $\nu'=-\sin 2\theta \nu_1+\cos 2\theta \nu_2$.  
We denote by $P$ the corresponding two-neutrino conversion probability. Taking the maximal mixing $\Theta=\pi/4$  in the matrix (\ref{eq:10}) 
we find the conversion matrix: 
\begin{eqnarray}  
{\cal P}=\pmatrix{ 1-P  & {P/ 2}  &  {P/ 2}  \cr
           {P/ 2}     & 1/2-{P/ 4}   
	   & 1/2-{P/ 4} \cr
	   {P/ 2}  	& 1/2-{P/ 4}    &  
	   1/2-{P/ 4}  \cr}~,   
\label{eq:pmatrix}
\end{eqnarray}
and an analogous expression for the matrix of probabilities for antineutrinos with the replacement $P\rightarrow \bar{P}$, where $\bar{P}$ represents the  $\bar{\nu}_e \rightarrow \bar{\nu'}$ conversion probability.

Taking  the CP-symmetric flavour composition (\ref{eq:cpsflux}), from eqs. (\ref{eq:pmatrix}) and (\ref{eq:matrixform}) 
we find  that  the conversion 
probability $P$ cancels in the expression of the numbers of events, $N_{\alpha}$. Equal numbers of events for the three flavours are predicted independently of matter
effects: $\vec{N}_\nu=\vec{N}_{\bar{\nu}}=(1,1,1)$.
 
For the CP-asymmetric composition (\ref{eq:cpaflux}) we obtain:
\begin{eqnarray} 
&&\vec{N}_\nu=\left(1-{P/ 2} , 1/2+{P/ 4}, 
1/2+{P/ 4}\right) \nonumber\\ 
&&\vec{N}_{\bar{\nu}}=\left({\bar{P}/ 2} ,
1/2-{\bar{P}/ 4}, 
1/2-{\bar{P}/ 4}\right)~. 
\label{eq:fluxes}
\end{eqnarray}
Since the present detectors do not distinguish neutrinos from
antineutrinos, we consider the sums of the events induced by $\nu$ and $\bar{\nu}$. From eqs. (\ref{eq:fluxes}) we find:
\begin{eqnarray} 
\vec{N}_\nu+ \vec{N}_{\bar{\nu}}=\left(1-{P/ 2}+{\bar{P}/ 2} , 1+{P/ 4}-{\bar{P}/ 4}, 1 +{P/ 4} -{\bar{P}/ 4}\right)~.
\label{eq:fluxessum}
\end{eqnarray}
Two comments are in order.
First, equal numbers of events induced by the muon and tau neutrinos
are expected, with no dependence of ratios on matter effects: 
$(N_{\mu}+N_{\bar{\mu}})/(N_{\tau}+N_{\bar{\tau}})=1$. Conversely, matter effects are present in ratios involving the electron neutrino.
Second, if $P=\bar{P}$ the conversion probability cancels in (\ref{eq:fluxessum}) and one gets $\vec{N}_\nu+ \vec{N}_{\bar{\nu}}=(1,1,1)$. 
This circumstance is realized in absence of matter effects ($F\eta=0$) or in the extremely high energy limit,  $E_0/\Delta m^2\gta 10^{33}~{\rm eV^{-1}}$, in which the asymptotic value (\ref{eq:asympprob}) for the conversion probability is realized (see also fig. \ref{fig:fig8new}).  Therefore, the matter effect could be revealed by a deviation from the equality  of number of events for the three flavours in the narrow interval $E_0/\Delta m^2\simeq 10^{32}-10^{33}~{\rm eV^{-1}}$ in which $P$ and $\bar{P}$ are unequal and have significant deviation from the vacuum oscillation probability.

Considering, for instance,
the ratio of e-like over non e-like events we find:   
\begin{eqnarray} 
R\equiv {N_{e}+N_{\bar{e}} \over 
N_{\mu}+N_{\bar{\mu}}+N_{\tau}+N_{\bar{\tau}}}
={1- {P/ 2}+{\bar{P}/ 2}\over 2+ {P/ 2}-{\bar{P}/ 2}}~,
\label{eq:ratio}
\end{eqnarray}

The deviation of $R$  from its
value $R_p=1/2$ without oscillations
 is entirely due to matter effects and equals:
\begin{eqnarray} 
{R-R_{p} \over R_{p}}\simeq -{3\over 2}\Delta ~,
\label{eq:expand}
\end{eqnarray}
where  $\Delta \equiv (P-\bar{P})/2$.
The relative deviation (\ref{eq:expand}) amounts to $\sim 15\%$ for $\Delta \simeq 0.1$.  Similar conclusions are
obtained for  other ratios of numbers of events. 

Results are different if the mixing angle $\Theta$ in the matrix (\ref{eq:10})  is not maximal. In the extreme case  $\Theta=0$ the problem reduces to two-neutrino conversion. In the limit $E_0/\Delta m^2\gta 10^{33}~{\rm eV^{-1}}$, for both the compositions (\ref{eq:cpsflux}) and (\ref{eq:cpaflux}) we get:
\begin{eqnarray} 
{R-R_{p} \over R_{p}}\simeq {3\over 2}P ~,
\label{eq:expandbis}
\end{eqnarray}
where we considered $P\simeq \bar{P}$. Taking  $P \simeq 0.1$ the deviation (\ref{eq:expandbis}) equals $\sim 15\%$.

\subsection{Extension to four neutrinos}
\label{subsec:5.4}

An example of four neutrino scheme with sterile neutrino, $\nu_s$, was introduced in section \ref{subsec:5.1}: the sterile state is present in the two light mass eigenstates, $\nu_0$ and $\nu_1$, so that in the bases $\vec{\nu}_\alpha=(\nu_s,\nu_e,\nu_\mu,\nu_\tau)$, $\vec{\nu}_i=(\nu_0,\nu_1,\nu_2,\nu_3)$ the mixing matrix takes the form:
\begin{eqnarray}
U^0=\pmatrix{ {c_\phi}  & {s_\phi}  & {0} & {0} \cr
 {-c_\theta s_\phi}  & {c_\theta c_\phi}  & {-s_\theta} & {0} \cr
 {-s_\theta s_\phi/\sqrt{2}}  & {s_\theta c_\phi/\sqrt{2}}  & {c_\theta/\sqrt{2}} & {-1/\sqrt{2}} \cr
{-s_\theta s_\phi/\sqrt{2}}  & {s_\theta c_\phi/\sqrt{2}}  & {c_\theta/\sqrt{2}} & {1/\sqrt{2}} \cr 
}~, 
\label{eq:umatrix4}
\end{eqnarray}
where $c_\phi\equiv \cos \phi$, $s_\phi\equiv \sin \phi$,  $c_\theta\equiv \cos \theta$, $s_\theta\equiv \sin \theta$.  
The angle $\phi$ describes the mixing between $\nu_s$ and the state $\tilde{\nu}\equiv c_\theta \nu_e+s_\theta \nu_\mu /\sqrt{2}+s_\theta \nu_\tau /\sqrt{2}$. Analogously to the previous case, we consider  
$\Delta m^2_{1 0}\lta 10^{-7} ~{\rm eV^2}$ and all the other mass splittings to be much larger than this value, so that the interaction with the neutrino background affects the propagation of the $\nu_0-\nu_1$ system only. As a consequence, the matter effect modifies the angle  $\phi$ only; the  changes of $\theta$ are negligibly small.
Again, the dynamics of the four neutrino system is reduced to the evolution of the two states $\nu_s$ and $\tilde{\nu}$. Introducing the conversion probability $P\equiv P(\nu_s\rightarrow \tilde{\nu})$, we find the matrix of probabilities (see eq. (\ref{eq:matrixform})): 
\begin{eqnarray}  
{\cal P}=\pmatrix{
{1-P} & {c_\theta^2 P} & {s_\theta^2 P/2} & {s_\theta^2 P/2} \cr
{c_\theta^2 P} & {s_\theta^4+c_\theta^4(1-P)} & {s_\theta^2 c_\theta^2 (1-P/2)} & {s_\theta^2 c_\theta^2 (1-P/2)} \cr
{s_\theta^2 P/2} & {s_\theta^2 c_\theta^2 (1-P/2)} & {[1+c_\theta^4+s_\theta^4(1-P)]/4} & {[1+c_\theta^4+s_\theta^4(1-P)]/4} \cr
{s_\theta^2 P/2} & {s_\theta^2 c_\theta^2 (1-P/2)} & {[1+c_\theta^4+s_\theta^4(1-P)]/4} & {[1+c_\theta^4+s_\theta^4(1-P)]/4} \cr
}~.   
\label{eq:pmatrix4}
\end{eqnarray}

Taking the  CP-symmetric composition (\ref{eq:cpsflux}) and assuming that no sterile neutrinos are produced, $N^0_s=0$, from eq. (\ref{eq:pmatrix4}) and (\ref{eq:matrixform}) 
one gets  the numbers of events:
\begin{eqnarray} 
&&\vec{N}_\nu=\left(P , 1-c_\theta^2 P, 1-s_\theta^2 P/2, 1-s_\theta^2 P/2\right) \nonumber\\ 
&&\vec{N}_{\bar{\nu}}=\left(\bar{P} , 1-c_\theta^2 \bar{P}, 1-s_\theta^2 \bar{P}/2, 1-s_\theta^2 \bar{P}/2\right)~. 
\label{eq:CPsfluxes4}
\end{eqnarray}
As in the three neutrino case, we have $(N_{\mu}+N_{\bar{\mu}})/(N_{\tau}+N_{\bar{\tau}})=1$ independently on matter effects. Notice that in the total numbers of events $\vec{N}_\nu+\vec{N}_{\bar{\nu}}$ the conversion probabilities appear in the combination $P+\bar{P}$: since the matter effects have opposite signs for neutrinos and antineutrinos, they partially cancel in this quantity.

Introducing the deviation from the averaged vacuum oscillation  probability, $\delta_P\equiv P+\bar{P}-2P_v$, we compute  the ratio:  
\begin{eqnarray} 
R\equiv {N_{e}+N_{\bar{e}} \over 
N_{\mu}+N_{\bar{\mu}}+N_{\tau}+N_{\bar{\tau}}}=
{ 1-c_\theta^2 (P_v+ \delta_P/2) \over 2-s_\theta^2(P_v+ \delta_P/2)}~.
\label{eq:cpsratio4}
\end{eqnarray}
The relative deviation of this ratio from the value given by vacuum oscillations equals:
\begin{eqnarray} 
{R-R_{v} \over R_{v}}\simeq -{\delta_P \over 2}{ c_\theta^2- s_\theta^2/2  
\over (1-s_\theta^2 P_v/2)(1-c_\theta^2 P_v)}  ~.
\label{eq:cpsexpand4}
\end{eqnarray}
Taking $\delta_P\simeq 0.1$, $s_\theta^2\simeq c_\theta^2\simeq 1/2$ and $P_v\simeq 0.4$, eq. (\ref{eq:cpsexpand4}) gives a deviation of $\sim 2\%$; the effect is larger, $\sim 10\%$, for small  $\theta$: $c_\theta^2\simeq 1$, $s_\theta^2\simeq 0$.  

For the CP-asymmetric composition (\ref{eq:cpaflux}) we get:
\begin{eqnarray} 
\vec{N}^0_\nu&=&\left(P (c_\theta^2+s_\theta^2 /2) , 1-s_\theta^2 c_\theta^2 -c_\theta^2 P(c_\theta^2+s_\theta^2 /2) \right. , \nonumber\\
&&\left. ( 1+s_\theta^2 c_\theta^2)/2-s_\theta^2 P(c_\theta^2+s_\theta^2 /2)/2, ( 1+s_\theta^2 c_\theta^2)/2-s_\theta^2 P(c_\theta^2+s_\theta^2 /2)/2\right) \nonumber\\ 
\vec{N}^0_{\bar{\nu}}&=&\left({s_\theta^2 \bar{P}/2} , {s_\theta^2 c_\theta^2 (1-\bar{P}/2)} , \right. \nonumber \\
&&\left. {[1+c_\theta^4+s_\theta^4(1-\bar{P})]/4} , {[1+c_\theta^4+s_\theta^4(1-\bar{P})]/4} \right)~. 
\label{eq:CPafluxes4}
\end{eqnarray}
For the ratio, $R$, of the e-like over non e-like events one gets:
\begin{eqnarray} 
{R-R_{v} \over R_{v}}\simeq -\left[\delta (1-s^2_\theta/2)+ {\bar{\delta}} s_\theta^2/2 \right] \left[ { c_\theta^2- s_\theta^2/2  
\over (1-s_\theta^2 P_v/2)(1-c_\theta^2 P_v)} \right]~,
\label{eq:cpaexpand4}
\end{eqnarray}
where $\delta\equiv P-P_v$ and $\bar{\delta} \equiv \bar{P}-P_v$. 

with the values $\delta\simeq 0.1$, $\bar{\delta}\simeq -0.05$,  $P_v\simeq 0.4$  and small mixing, $c_\theta^2\simeq 1$, $s_\theta^2\simeq 0$, the deviation (\ref{eq:cpaexpand4}) equals $\sim 15\%$ similarly to the case of CP-symmetric composition, eq. (\ref{eq:cpsexpand4}). The effect is smaller, $\sim 2\%$, for large mixing, $s_\theta^2\simeq c_\theta^2\simeq 1/2$. 
\\

Our estimation, $10-15\%$ effect, gives some hope that the discussed phenomenon will be observed in future large scale experiments with event rates $\sim 1000$ events/year.

\section{Conclusions}
\label{sec:5}

We have studied matter effects on oscillations of high energy cosmic neutrinos.
The only known component of the intergalactic  medium which can contribute to such an effect is the relic neutrino background provided that it has large CP (lepton) asymmetry. 
\\

\noindent
The mixing modifies the flavour composition of the relic neutrino background. 
Considering atmospheric and solar neutrino-motivated mixings and mass squared differences we find that, if large  asymmetries in the muon and/or tau flavours are produced before the BBN epoch, they are equilibrated by the combined effect of oscillations and inelastic collisions, so that $\eta_\mu\simeq \eta_\tau$. The asymmetry in the electron flavour, $\eta_e$, can be equilibrated with $\eta_\mu$ and $\eta_\tau$ for $\Delta m^2\gta 10^{-7}~{\rm eV^2}$.  For $\Delta m^2\lta 10^{-7}~{\rm eV^2}$ $\nu_e-\nu_\mu/\nu_\tau$ oscillations are suppressed by collisions and/or by the expansion rate of the universe, thus leaving $\eta_e$ unchanged, at least until the BBN epoch.
At later epochs oscillations develop and   large asymmetries in the muon and/or tau flavours can be efficiently converted into $\nu_e$ asymmetry.  Therefore  at present the values of the asymmetries for the three flavours can be comparable.  This allows one to reconcile possible large lepton asymmetry in the $\nu_e$ flavour at present, $\eta_e\sim 1$, with strong constraint on $\eta_e$ from nucleosynthesis.  Active-sterile conversion is ineffective until the BBN or later, due  the matter-induced suppression of the $\nu_\alpha-\nu_s$ mixing.
\\

\noindent
The dynamics of high-energy neutrino conversion in the CP-asymmetric neutrino background has been considered. For conversion between active neutrinos the matter effects consist in a modification of the vacuum oscillation length. 
The effect is significant for large mixing angle, $\sin^2 2\theta\gta 0.3$, and high energies, ${E_0 / \Delta m^2}\gta  5 \cdot 10^{32} ~{\rm eV^{-1}}$, for which the matter contribution to the oscillation phase dominates over the vacuum oscillation one.  In these circumstances the conversion probability can differ by $\sim 30\%$ from the vacuum oscillations value.
\\

\noindent
For active-sterile conversion the matter effects can be important in the interval ${E_0 / \Delta m^2}\gta  10^{28}-10^{32} ~{\rm eV^{-1}}$, for which the resonance condition is satisfied.
For the majority of realistic situations (with $z\lta 10$, $\eta\lta 10$), the adiabaticity condition is broken.  This implies that the matter effect is reduced to non-adiabatic level crossing or enhancement (suppression) of mixing and therefore of the depth of oscillations in the production epoch.   The relative change of the conversion probability can be as large as $20-50\%$.  For extreme values of the asymmetry and large redshift of production the adiabatic conversion can take place with almost maximal conversion probability.
\\

\noindent
We calculated the effect of conversion on the diffuse fluxes of neutrinos produced by GRBs, AGN and the decay of super-heavy relics.
For neutrinos from GRBs and AGN the relative deviation of the flux due to matter effects with respect to vacuum oscillations can reach $20\%$.   
For neutrinos from heavy particle decay the effect can be larger: up to $\sim 40\%$.
\\

\noindent
Possible signatures of matter effects consist in the deviation of ratios of numbers of observed events,  $N_{e}/N_{\mu}, N_{e}/N_{\tau}, N_{\mu}/N_{\tau}$, from the values predicted by pure vacuum oscillations.  Presumably, neutrino mixings and masses will be measured in laboratory experiments and vacuum  oscillations  effects will be reliably predicted.

For conversion into a sterile state one expects also a characteristic energy dependence of the ratios which in principle will allow to distinguish matter effects from the uncertainties in the flavour content of original neutrino fluxes.  

For illustration purpose we estimated observable effects for two possible schemes of neutrino masses and two different flavour compositions of the detected
 fluxes in absence of conversion.  In a scheme with three flavour states only and parameters in the region of  VO solution of the solar neutrino problem we found that the deviation of ratios of numbers of events from their vacuum oscillation values can be of  $\sim 10\%$.   Similar conclusion is obtained for schemes with an additional sterile neutrino.
\\

\noindent
Clearly, more work is needed to clarify the possibilities to observe the effect under consideration.  In any case, new large scale detectors with relatively high statistics  ($\sim 1000 ~{\rm events/year}$) are required.

The detection of matter effects on fluxes of high-energy neutrinos would be an evidence of  large CP (lepton)-asymmetry in the universe.  As follows from our analysis, asymmetries of order $\eta\sim 1$ can be probed in these studies. 
Clearly, the observation of such a large asymmetry will have far going consequences for our understanding of the evolution of the universe.


\bibliography{eunew}
\end{document}